\documentstyle[preprint,prb,aps]{revtex}
\def\sigmav{{\mbox{\boldmath{$\sigma$}}}}
\def\rhov{{\mbox{\boldmath{$\rho$}}}}
\def\kappav{{\mbox{\boldmath{$\kappa$}}}}

\begin{document}
\input{psfig}
\title{ROTATIONAL AND VIBRATIONAL DYNAMICS OF INTERSTITIAL MOLECULAR HYDROGEN}
\author{T. Yildirim}
\address{National Institute of Standards and Technology,
Gaithersburg, Md 20899}
\author{A. B. Harris}
\address{Department of Physics and Astronomy, University of Pennsylvania,
Philadelphia, PA 19104}
\address{and National Institute of Standards and Technology,
Gaithersburg, Md 20899}
\date{\today}
\maketitle

\begin{abstract}
The calculation of the hindered roton-phonon energy levels 
of a hydrogen molecule in a confining potential with different
symmetries is systematized for the case when the rotational
angular momentum $J$ is a good quantum number. One goal of
this program is to interpret the energy-resolved 
neutron time of flight spectrum previously obtained for H$_{2}$C$_{60}$.  
This spectrum gives  direct information on the energy level
spectrum of H$_2$ molecules confined to the octahedral interstitial
sites of solid C$_{60}$.  We treat this problem of coupled translational
and orientational degrees of freedom a) by construction of
an effective Hamiltonian to describe the splitting of the manifold
of states characterized by a given value of $J$ and having a fixed
total number of phonon excitations, b) by numerical solutions of
the coupled translation-rotation problem on a discrete mesh of points
in position space, and c) by a group theoretical symmetry analysis.
Results obtained from these three different approaches are mutually
consistent.  The results of our calculations explain several hitherto
uninterpreted aspects of
the experimental observations, but show that a truly satisfactory
orientational potential for the interaction of an H$_2$ molecule with
a surrounding array of C atoms has not yet been developed.
\end{abstract}
\pacs{78.70.Nx,34.50.Ez,82.80.Gk,71.20.Tx,36.20.Ng,63.22.+m}

\section{INTRODUCTION}

The study of rotational and vibrational dynamics of 
guest molecules (i.e. CO, O$_{2}$,  H$_{2}$, etc)  trapped in porous
media such as fullerenes, zeolites, and graphite
has recently become an active subject both experimentally
and theoretically.\cite{gs1,gs2,gs3,novaco,NIST}
This is because such studies can
yield valuable  information about the host-guest interactions which could
be important for several technical applications such as
gas separation and hydrogen storage.\cite{gs1,gs2,gs3} In particular 
hydrogen molecules trapped  in interstitial cavities in
solid C$_{60}$ as well as hydrogen molecules embedded in
nanotube ropes are of interest due to quantum behavior  of
hydrogen molecules in quasi zero and one dimensional 
sites.\cite{gs2,gs3,NIST}

In this paper, we develop a detailed analysis of coupled 
rotational and vibrational dynamics  of a molecular hydrogen
encapsulated in a solid using numerical, perturbative, and
group theoretical methods. 
In particular we will be interested in
what one might call the ``weak coupling limit,'' when
the interaction between molecular rotations and center-of-mass
translations is weak enough that the rotational angular momentum
quantum number $J$ is a good quantum number.  This limit is
almost never satisfied except for very light molecules like
hydrogen or deuterium.  The energy levels of a free rotator
are
\begin{eqnarray}
E_J = B J (J+1) \ ,
\end{eqnarray}
where $B = \hbar^2/(2I)$, where $I$ is the moment of inertia of the
molecule and $E_J$ is $(2J+1)$-fold degenerate.  For H$_2$
the rotational constant $2B$ has the value $60$ cm$^{-1}$, 14.7 meV,
or $B/k=85$ K (and the corresponding values for D$_2$ are half as large),
so that the energy separation between different $J$ levels is large
enough that often $J$ is a good quantum number.  This is
certainly true for solids consisting of these molecules unless the
pressure is quite large.  (For a review of the properties of the
hydrogen molecule and solid hydrogen see Ref. \onlinecite{VK}.)

We have been led to consider this phenomenon in view of an
experimental study of energy spectra of H$_2$ and D$_2$ 
inserted into the octahedral interstitial sites in solid
C$_{60}$ carried out by neutron time of flight techniques.\cite{NIST}
In considering this phenomenon we should keep in mind the following
experimental facts concerning the host solid of C$_{60}$.
The centers of the C$_{60}$ molecules form  an fcc lattice.\cite{FM3M}
At temperature above $T_c$, where $T_c$ is about 260~K, the molecules
are orientationally disordered. At $T=T_c$ long range orientational
ordering occurs\cite{PAH} and the molecules are ordered into
four sublattices as described by Pa$\overline 3$ symmetry.\cite{pa1,pa2,pa3,pa4}
In the orientationally disordered phase the local symmetry at the
the octahedral interstitial site is indeed that of the point group O$_h$.
In the presence of orientational ordering the symmetry of what was the
``octahedral'' interstitial site is now reduced to a uniaxial symmetry,
specifically that of point group S$_6$.\cite{HAHN} In experiments,
hydrogen molecules are stable in the octahedral interstitial site
only for temperatures well below $T_c$ (where the interstitial site
does not actually have octahedral symmetry).  

While a general understanding of the time-of-flight experiments
was presented,\cite{NIST}
some of the finer details of the experiment remained unexplained.
For instance, the shift in the energy associated with ortho-para
($J=1 \rightarrow J=0$) conversion in the interstitial relative
to its value for free molecules was not understood.  Also
the feature in the energy gain
spectrum at about twice the ortho-para conversion energy was
not unambiguously identified.  These issues are both
addressed in this paper.  More generally we give a calculation
of the cross section for neutron energy loss for comparison with 
the observed time-of-flight spectrum.  For that purpose we
need not only to consider the cross section for para-ortho
conversion as compared to phonon creation, but also
to calculate the phonon excitations of $(J=1)$ molecules.
These calculations require us to develop and implement a
scheme for treating coupled translations and rotations. 
In this paper we present a systematic analysis of the simplest
case of this coupling which occurs when the quantum number
$J$ characterizing free rotation remains a good quantum number.
In that case the well-known numerical schemes for solving
the translation problem can be easily extended to include
the effect of the coupling to rotations.  In addition,
we also give analytic expressions obtained by treating
this coupling within perturbation theory.  As we will see,
this analytic development enables us to interpret many of
the numerical results in a meaningful way.  In addition,
we analyze in detail various simplified models which
illustrate our group theoretical analysis of the
symmetry present in the system of coupled translations
and rotations.  This analysis indicates that arguments for
the degeneracies of coupled translation-rotation modes based
on simple classical concepts are incorrect.  In summary:
in this paper we present an analysis based on numerical,
perturbative, and group theoretical methods.

Up to now there have not been many theoretical studies of energy levels
in such irregular geometries like the octahedral interstitial sites
in C$_{60}$.  A notable exception is the work of van der Avoird and
collaborators\cite{AVOIR}
on CO in C$_{60}$.  That work examined an even more complicated situation
in which the rotational and translation degrees of freedom 
interacted strongly.  As a result, the problem was analyzed
numerically.  In contrast, for the present problem
FitzGerald {\it et al.}\cite{NIST}
applied a number of analytic and semi-analytic techniques to the
theoretical study the spectra of hydrogen molecules in C$_{60}$.
This paper may be regarded as an extension and systematization
of their approach.

\section{GENERAL FORMULATION}

Clearly the first step is to establish a satisfactory potential
for the intercalated hydrogen molecule.  This potential function
$V({\bf r},\Omega)$ gives the energy of a hydrogen molecule whose
center of mass is at $\bf r$ and whose orientation is specified by
$\Omega \equiv (\theta , \phi)$.  A convenient starting point is
to use an atom-atom potential\cite{ATOM} to describe the interaction
between each of the two hydrogen atoms and the atoms in the
confining structure.  Unless otherwise indicated, 
all the results reported in this paper
are obtained from the same WS77 potential,\cite{ATOM} 
$ -A/r^{6}+B \exp(-Cr)$, that
is used in Ref.\onlinecite{NIST} (where 
$A=5.94\ {\rm eV} \AA^{6}, B= 678.2$ eV, and $C=3.67 \AA^{-1}$).

In this paper we will mainly consider
the octahedral interstitial site in solid C$_{60}$, but many of
the considerations apply with slight modification to molecules
confined within other structures such as
single wall carbon nanotubes.\cite{YHII} The determination of
the potential $V({\bf r},\Omega)$ for H$_2$ in solid C$_{60}$
is discussed in Appendix \ref{YLM}.  From the numerical
evaluation of this potential we have extracted the expansion
coefficients when it is written in the following canonical form:
\begin{eqnarray}
\label{AEQ}
V({\bf r}, \Omega ) = V_0 ({\bf r}) +
\sum_{l=2,4, \dots} \sum_{m = -l}^l A_l^m ({\bf r}) Y_l^m(\Omega) \ .
\end{eqnarray}
We assume (and it is generally true) that the orientational
energies which are relevant are much less than the
smallest energy difference between successive $J$ levels of a
molecule ($10B$ for an
ortho molecule and $6B$ for a para molecule).  Accordingly,
we may consider only that part of the potential which is diagonal
in $J$.  When the potential is written in the form of Eq. (\ref{AEQ}),
it is easy to implement the truncation to terms diagonal in $J$.
So for a fixed value of $J$ we have the Hamiltonian ${\cal H}_J$ as
\begin{eqnarray}
{\cal H}_J & = & {p^2 \over 2m} + V_0 ({\bf r}) + BJ(J+1) {\cal I}
+ \sum_{l=2,4, \dots} \sum_{\mu, \mu'} A_l^{\mu-\mu'}({\bf r}) \Biggl[
| J \mu \rangle \langle J \mu| Y_l^{\mu-\mu'}(\Omega) | J \mu' \rangle
\langle J \mu' | \Biggr] \nonumber \\
&=& - {\hbar^2 \over 2m} \nabla^2 + V_0 ({\bf r}) + BJ(J+1) + V_J({\bf r}) \ ,
\label{HJEQ} \end{eqnarray}
where $V_J({\bf r})$ is the orientationally dependent part of the potential
(the terms involving $A_l^m$) and ${\cal I}$ is the unit operator.
Furthermore we consider the angular-dependent term in this expansion
to be a perturbation on the first term, $V_0({\bf r})$.
For each value of $J$ the Hamiltonian ${\cal H}_J$
will give a manifold of states which is the
direct product of a manifold corresponding
to various numbers of localized phonons being excited with the manifold
of $(2J+1)$ states having different values of $m_J$.  An important
simplification is that spherical harmonics with $l > 2J$ have no
nonzero matrix elements in the manifold of states of angular momentum $J$.

Note that apart from the kinetic energy, this Hamiltonian is a
strictly local operator.  Thus we solve the eigenvalue problem on
a discrete mesh of points on a cube centered at the octahedral site
when the wave function is required to vanish on the boundary of
the cube. Each edge of the cube is taken to be
$[-L,L]$ with mesh point spacing of $dL$. 
In this scheme the wave function at
each mesh point is a $(2J+1)$-component vector.  
We are mainly concerned
with the manifold $J=0$ and $J=1$, in which case the problem is numerically
not significantly harder than for a scalar problem. Even though the
resulting matrix size is very large, it is a block band matrix and
is very sparse. 
The numerical results
reported here were obtained from L=1.65 \AA \ and $dL=0.075 $\ \AA,
which requires diagonalization of a matrix $n\times n$ where
$n=273375$). However we confirmed that a coarse mesh points
with $L=1.2$ \AA \ and $dL=0.17$ \AA, gives almost the same
results (where the matrix size is $n=14,739$). The large sparse matrix
eigenvalue problem is solved using the package ARPACK.\cite{ARPACK}

Since numerical results sometimes do not provide complete insight into
the nature of the solutions, we have also used perturbation theory to
understand the results.  In  this approach we treat $V_J({\bf r})$ in
Eq. (\ref{HJEQ}) as the perturbation.  The unperturbed problem, apart
from the additive energy $BJ(J+1)$ is thus that for translations of
the spherical $(J=0)$ molecule.  This spectrum is not too different from
that of a three-dimensional harmonic oscillator.  Accordingly, to
qualitatively interpret our more accurate numerical results, we apply
perturbation theory in which we develop an effective Hamiltonian\cite{KITTEL}
to describe the splitting of this manifold which is characterized by a
value of $J$ and of $N$, the total number of phonon excitations.
(The perturbative effects due to coupling between manifolds of
different $J$ is negligibly small for hydrogen in C$_{60}$.\cite{NIST})
This effective Hamiltonian is a matrix of dimensionality $D$, where
$D=(2J+1)(N+1)(N+2)/2$ and schematically is of the form
\begin{eqnarray}
{\cal H}(N,J) &=& BJ(J+1){\cal I} + {\cal H}_{\rm phonon} + V_J^{\rm DIAG}
- V_J^{\rm OFF} {1 \over {\cal E}} V_J^{\rm OFF} \ ,
\end{eqnarray}
where ${\cal H}_{\rm phonon}$ gives the energy of the various states
with a total of $N$ phonons.  These energies are just those calculated for
a $(J=0)$ molecule.  Also $V_J^{\rm DIAG}$ is the part of $V_J$ which
is diagonal {\it with respect to the number of phonons},
$V_J^{\rm OFF}$ is the part of $V_J$ which is off-diagonal
{\it with respect to the number of phonons}, and
${\cal E}$ is the change in phonon energy caused by 
$V_J^{\rm OFF}$.

This effective Hamiltonian is defined by its matrix elements as
\begin{eqnarray}
\label{HEFFEQ}
\langle N, \alpha &;& J, M | {\cal H}(J,N)
| N, \alpha' ; J, M' \rangle =
[BJ(J+1) + E_{N,\alpha} ] \delta_{\alpha, \alpha'} \delta_{M,M'}
\nonumber \\ && \ + \sum_{l=1}^J
\langle N, \alpha |A_{2l}^{M-M'} ({\bf r})| N, \alpha' \rangle
\langle J M |  Y_{2l}^{M-M'}(\Omega) | J M' \rangle \nonumber \\
&& + \sum_{N' \not= N} \sum_{l,l'=1}^J \sum_\mu \sum_{\beta =1}^{k_{N'}}
[ E_{N, \alpha} - E_{N', \beta} ]^{-1}
\langle N, \alpha| A_{2l}^{M-\mu}({\bf r}) | N' , \beta \rangle
\langle N' , \beta | A_{2l'}^{\mu-M'} ({\bf r})| N, \alpha' \rangle
\nonumber \\ && \ \
\langle J M |  Y_{2l}^{M-\mu}(\Omega) | J \mu \rangle
\langle J \mu | Y_{2l'}^{\mu-M'} (\Omega) | J M' \rangle \ ,
\end{eqnarray}
where $k_N=(N+1)(N+2)/2$ and the states with $N$ phonons are labeled
$N,\alpha$, where $\alpha$ runs from 1 to $k_N$.

We now briefly discuss how the $A_l^m$'s of Eq. (\ref{AEQ})
are obtained from the atom-atom
potential between each H atom and each carbon atom.  Here we will
assume that the atom-atom potential is of the form 
$F(|{\bf r}_i - {\bf r}_H|)$,
where ${\bf r}_i$ and ${\bf r}_H$ are the displacements of the $i$th 
carbon and of the
H atom, respectively, relative to the center of the H$_2$ molecule.
For this form of potential, we show in Appendix
\ref{YLM} that
\begin{eqnarray}
A_2^m &=& 2 \pi \sum_i Y_2^m(\hat {\bf r}_i)^*  \int_{-1}^1 (3x^2-1)
F \left( [r_i^2 - \case 1/4 \rho^2 + \rho x r_i]^{1/2} \right) dx \ ,
\label{BUILDA} \end{eqnarray}
where $\rho$ is the separation between H atoms in the H$_2$ molecule and the
sum over $i$ is over all relevant neighboring carbon atoms. It is
instructive to expand this expression in $\rho/r_{i}$, which yields
\begin{equation}
A_2^m =  \frac{1}{4} \rho^{2} \biggl ( \frac{8\pi}{15}  \biggr)
\sum_i \biggl ( F^{''} - \frac{F^{'}}{r_{i}} \biggr ) Y_2^m(\hat {\bf r}_i) + 
o(\rho^{4}/r_{i}^{4}),
\end{equation}
where $F^{''}$ and $F^{'}$ are the second and first derivatives of
$F(r_{i})$.

This expansion is good enough to reproduce most of the results discussed
in this paper. We note that the $A_l^m$'s (i.e. the orientational potential)
are zero for a harmonic potential (i.e.  $F(r_{i})= \frac{1}{2} k r_{i}^{2}$)
because the prefactor $( F^{''} - F^{'}/r_{i} \biggr )$
is zero. This can be also seen easily as follows.
Assuming  an atom-atom potential between each H atom and the C atoms
in the adjacent C$_{60}$ molecules, we may write the potential of
an H$_2$ molecule as
\begin{eqnarray}
V({\bf r}; \Omega) &=& V_a ({\bf r} + \case 1/2 \rho \hat n)
+ V_a ({\bf r} - \case 1/2 \rho \hat n) \ ,
\label{AAEQ} \end{eqnarray}
where $V_a$ is the potential of a single atom due to the entire
octahedral cage in which it is confined, $\hat n$ is a unit vector
along the axis of the molecule,
and $\rho$ is the separation between atoms in the molecule.  As we shall see,
the total potential is nearly isotropic.  So we write
\begin{eqnarray}
V_a({\bf r})  &=& \case 1/2 k r^2 + \delta r^4 \ .
\end{eqnarray}
When we substitute this into Eq. (\ref{AAEQ}), we obtain the result
\begin{eqnarray}
V({\bf r}; \Omega )&=&  V_0(r)
+ 2 \delta [r^2 \rho^2 \cos^2 \theta_{r,n} - \case 1/3 ] \ ,
\end{eqnarray}
where $\theta_{r,n}$ is the angle between the vectors ${\bf r}$ and $\hat n$
and $V_0$ is independent of $\theta_{r,n}$.
The point is that the orientationally dependent part of the interaction depends
on the anharmonicity: for a purely harmonic and isotropic interaction $V_a$,
the total potential energy is independent of the molecular orientation.  Thus
we expect rotation-translation coupling to be weak.  On the other hand in
nanotubes, where the quadratic term is
{\it anisotropic}, this coupling will be more important.\cite{YHII}

\section{ENERGY SPECTRUM OF A $(J=0)$ H$_2$ MOLECULE}

We start by considering the eigenvalue spectrum of ${\cal H}_0$
in which the orientational dependence of the potential is neglected.
In this approximation, apart from the additive constant $BJ(J+1)$,
the total energy (rotational plus translational) is the same as that of
a $(J=0)$ molecule.  For most purposes a $(J=0)$ molecule
may be considered to be a spherical molecule because the orientational
wave function $Y_0^0(\Omega)$ is uniform over all orientations.
Each eigenfunction of ${\cal H}_0(J)$ is the product of a rotational
function taken from the manifold of $2J+1$ degenerate orientational
wave functions and a translational wave function which represents
an eigenfunction for a spherical molecule confined to a cage.
These translational wave functions satisfy
\begin{eqnarray}
\label{JZEQ}
{\cal H}_0 \psi_k ({\bf r}) =
\Biggl[ {p^2 \over 2m} + V_0 ({\bf r}) \Biggr] \psi_k({\bf r})
= E_k \psi_k ({\bf r} ) \ ,
\end{eqnarray}
where $V_0({\bf r})$ is the potential discussed in Appendix \ref{YLM}.
The index $k$ labels states which we might otherwise label by three
indices, each quantum number characterizing the number of excitations
in each direction.  Note that these unperturbed  solutions do {\it not}
involve the coupling between rotations and translations.  As
discussed above, these eigenfunctions were
obtained by converting the continuum equation (\ref{JZEQ}) into a
into a discrete equation on a mesh of points and solving
the resulting matrix eigenvalue problem using a sparse matrix
routine.\cite{ARPACK}

\begin{figure}
\vspace{0.2 in}
\centerline{\psfig{figure=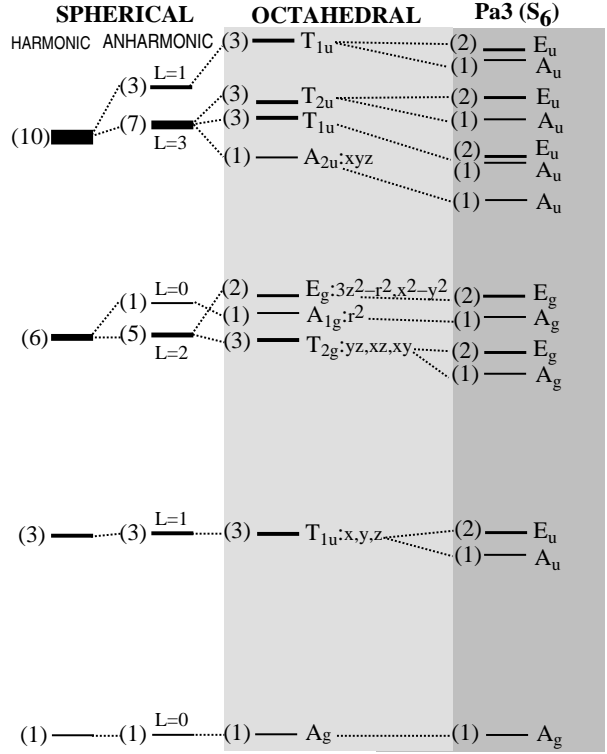,width=8cm}}

\vspace{0.2 in}
\caption{Energy levels of a spherical $(J=0)$ H$_2$ molecule confined
in various ways.  Here $(g)$ is the degeneracy and the symmetry labels
are given.  Left: The molecule is in a spherical harmonic
potential $V({\bf r})= \case 1/2 k(x^2+y^2+z^2)$ or an anharmonic
spherically symmetric potential (i. e. a generic spherically
symmetric potential). 
For a harmonic spherically symmetric potential the energy depends only
on $N$, the total number of phonon excitations in the oscillators along
the three coordinate directions.  For a spherically symmetric potential
eigenstates are characterized by their total orbital 
angular momentum $L$.
Center: the
molecule is in a potential appropriate to the octahedral interstitial
site of orientationally disordered (Fm3m) solid C$_{60}$.
Right: the
molecule is in a potential appropriate to the octahedral interstitial
site of orientationally ordered (Pa${\bar 3}$) solid C$_{60}$, in which case
the site symmetry is S$_6$. The potentials used for the
interstitial cases are discussed in Appendix \ref{YLM}.}
\label{PARAENERGY} \end{figure}

Since it happens that the energy levels  and eigenfunctions we obtained
numerically are not qualitatively different from those of a spherical
harmonic oscillator, we first study the energy spectrum 
as perturbations, $\delta, \kappa$, and $\lambda$,
are sequentially turned on in the following potential:
\begin{equation}
V(r) = \frac{1}{2} k r^{2} + \delta r^{4} + 
\kappa( x^4+y^4 + z^4 - \case 3/5 r^4)
+ \lambda(xy+yz+zx).
\end{equation} 
Fig. \ref{PARAENERGY} shows the evolution of the energy spectrum as
perturbations are sequentially introduced which take the spherical
harmonic oscillator into the
actual lower symmetry of a molecule in an octahedral interstitial site.
In the left-most panel we show the energy levels
for a spherical harmonic oscillator, with $\hbar \omega$ adjusted to
correspond to the single-phonon levels of H$_2$ in an
octahedral interstitial site in C$_{60}$.  Note that the levels
are highly degenerate because the energy depends only on the
total number of quanta.  The symmetry of the Hamiltonian is $U(3)$,
the group of unitary three dimensional matrices.
We now add to this potential an anharmonic term of the form
$\delta r^4$.  This perturbation lowers the symmetry to
that of the rotation group in three dimensions.  As is well known,
each eigenfunction in a generic spherically symmetric potential can
be labeled by the magnitude of the orbital angular momentum, $L$.  Thus the
single phonon levels are unsplit by this anharmonic perturbation and
are now labeled as angular momentum $L=1$ states, whereas the two phonon
levels split into a manifold of five $L=2$ states and one $L=0$ state
and similarly for states with more than two phonons.  In Fig.
\ref{PARAENERGY} we have taken the constant $\delta$ to be that which
best describes the anharmonicity of H$_2$ in C$_{60}$.  The
energies of the perturbed levels are given in Table \ref{AHTAB}.

Next, we consider what happens when the spherical oscillator potential
is augmented by a cubic symmetry potential of the form
$\kappa( x^4+y^4 + z^4 - \case 3/5 r^4)$.  This potential is appropriate
for a spherical molecule in an octahedral
interstitial when the C$_{60}$ are orientationally disordered
and have an Fm3m crystal structure.\cite{FM3M} 
The degeneracy associated with spherical symmetry
is lifted,\cite{GROUP} but as shown here one retains
cubic symmetry, so the three one-phonon states which transform as
$x$, $y$, and $z$ are degenerate.  The two-phonon states are of
three different symmetries.  One ($t_{2g}$) transforms like $xy$,
$xz$, and $yz$.  This is the lowest level.  The next highest level
is the s-wave symmetric combination which transforms like $x^2+y^2+z^2$.
Then one has a doublet of d-wave ($e_{2g}$) symmetry.  This
classification scheme is continued in the higher-energy levels.
Although we are not dealing with harmonic phonons,
it is still useful to consider manifolds
characterized by the quantum numbers $J$ and $N$, which are
respectively the rotational angular momentum and the total
number of phonons, at least up to $N=3$. Quantitative results
are given in  Table \ref{DEGENERACY}.

Finally, in the rightmost panel of Fig. \ref{PARAENERGY}
we show the further reduction
in degeneracy which occurs when the octahedral interstitial is
surrounded by C$_{60}$ molecules which have the long range order
associated with the Pa${\bar 3}$ crystal structure.\cite{pa1,pa2,pa3,pa4}
In this case, each interstitial is uniaxial (with symmetry S$_6$)
rather than octahedral.  Accordingly, we introduce a potential of the
form $\lambda(xy+yz+zx)\equiv \case 1/2 \lambda (3 \xi^2-r^2)$, where
the $\xi$ axis is taken to lie along the three-fold axis of the
interstitial site.  There are four symmetry related interstitial
sites, each of which has its three-fold axis along a
different $[1,1,1]$ direction.  The resolution of degeneracy in
the presence of this uniaxial perturbation
is also given in Table \ref{DEGENERACY}.  In all these cases, no
interactions between rotations and translations are involved.  

We have solved the eigenvalue problem of Eq. (\ref{JZEQ}) 
on a mesh of points and obtained the results given in Table
\ref{ENLEVEL}.  Results labeled "Octahedral" are those for
the orientationally disordered phase, where each C$_{60}$
molecule is replaced by a sphere of carbon atoms as is
discussed in Ref. \onlinecite{NIST}.  Since these numerical
results lead to manifolds of energy levels associated with
a given number of phonons and the degeneracies of these
manifolds are  as expected from
our general discussion above, we conclude that the potential
seen by a spherical H$_2$ molecule in the low-lying phonon levels
is not very different from that of a spherical harmonic oscillator.
However as noted in Ref.\onlinecite{NIST}, the effective
harmonic potential must be taken to be a self-consistently
renormalized potential to take account of the larger zero-point
motion.

\vspace{0.2 in}

\section{ENERGY SPECTRUM OF A $J=1$ MOLECULE}

We now discuss the energy spectrum of an ortho molecule with
$(J=1)$.  As we have seen for $(J=0)$ molecules, our
numerical results indicate that for $N$ up to, say, three, one
can clearly identify the manifold of $N$ phonons.  We therefore
discuss the systematics of these manifolds.

\subsection{Zero-Phonon Manifold}

We first consider the case of $J=1$ with $N=0$ phonons.
This manifold is described by the effective Hamiltonian
\begin{eqnarray}
{\cal H} &=& [2B + \Delta E ] {\cal I} + \delta [ J_z^2- \case 2/3 ] \ .
\end{eqnarray}
The splitting $\delta$ must be zero when C$_{60}$ is orientationally
disordered.  From Eq. (\ref{HEFFEQ}) one
sees that because the spherical harmonics are traceless,
the average energy shift, $\Delta E$, has nonzero contributions
only from terms which involve coupling to excited phonon states.
(In Ref. \onlinecite{NIST} a negligibly small shift was found due
to effects off-diagonal in $J$ which we ignore here.)
From Eq. (\ref{HEFFEQ}) we find that
\begin{eqnarray}
\Delta E &  = & - \case 1/3 \sum_{N \not = 0, \alpha} E_{N, \alpha}^{-1}
\sum_{\mu , \tau } | \langle 0,1| A_2^\tau ({\bf r}) | N,\alpha  \rangle |^2
| \langle 1 (\mu + \tau ) |  Y_2^\tau (\Omega) | 1 \mu \rangle |^2 \ .
\label{EQ12} \end{eqnarray}
To implement this equation, we first construct $A_l^m({\bf r})$ as
discussed in Eq. (\ref{BUILDA}).  Then matrix elements of
$A_2^M({\bf r})$ are taken between phonon states for a $J=0$
molecule which we obtained previously and  which are labeled $N,\alpha$
($|0,1 \rangle $ being the phonon ground state).  Thereby we obtained
the results given in Tables \ref{ALM0N} and \ref{A2MNNP}. 
In Eq. (\ref{EQ12}) the matrix elements of spherical harmonics
$Y_2^M (\Omega)$ are taken between orientational states
labeled by $J$ and $J_z$.  To evaluate $\Delta E$ we use
\begin{eqnarray}
\sum_\mu  | \langle 1 (\mu + \tau) |  Y_2^\tau (\Omega) | 1 \mu \rangle |^2
= {3 \over 10 \pi }
\label{DELEEQ}
\end{eqnarray}
so that
\begin{eqnarray}
\label{CGEQ}
\Delta E &  = & - {1 \over 10 \pi }
\sum_{N \not = 0, \alpha} E_{N, \alpha}^{-1}
\sum_\tau | \langle 0,1| A_2^\tau ({\bf r}) | N, \alpha \rangle |^2 \ .
\end{eqnarray}

In appendix \ref{CAVITY} we give a model calculation of an H$_2$ molecule
in a spherical cavity from which we evaluate Eq. (\ref{CGEQ}) to give
$\Delta E =  -0.14 {\rm meV}$.  In this calculation the translational wave
functions are assumed to be those of a harmonic oscillator with
$\langle r^2 \rangle = 0.1875 \AA^2$.  As noted, the result is very
sensitive to the value used for $\langle r^2 \rangle$.
For octahedral symmetry (i. e. for orientationally disordered C$_{60}$)
we evaluate Eq. (\ref{CGEQ}) using the data in Table \ref{ALM0N}. 
Thereby we find a shift $\Delta E = -0.133$ meV.  The same approach
using our numerical solutions for the phonon states of a
$(J=0)$ molecule
for the orientationally ordered Pa${\bar 3}$ phase yields the result,
$\Delta E = - 0.141 \ {\rm  meV} \ $,
compared to the experimental value\cite{NIST}
$\Delta E = - 0.35$ meV.  Again we mention that a small change in parameters
could easily lead to a much larger calculated value of $\Delta E$.
From the numerical solution for the three component wave function of
a $(J=1)$ molecule on a mesh of points, we obtained the value
$\Delta E = -0.16$ for the Pa${\bar 3}$ phase.  The various numerical results
for $\Delta E$ are summarized in Table \ref{SPLIT}.

From Eq. (\ref{HEFFEQ}) we also find the splitting (in the Pa${\bar 3}$ 
phase) to be
\begin{eqnarray}
\delta  &=& - {3\langle 0,1 | A_2^0 | 0,1 \rangle \over \sqrt{20 \pi} }
\nonumber \\ && \ + {3 \over 20 \pi} \sum_e E(e)^{-1} \Biggl[
|\langle 0,1 | A_2^0 ({\bf r}) | e \rangle |^2
+ |\langle 0,1 | A_2^1 ({\bf r}) | e \rangle |^2
- 2 |\langle 0,1 | A_2^2 ({\bf r}) | e \rangle |^2 \Biggr] \ , 
\label{SPLITEQ} \end{eqnarray}
where the quantization axis is taken to lie along the threefold axis
of symmetry of the interstitial site.
Using the matrix elements given in Table \ref{ALM0N}, we find that the
contribution to the splitting $\delta$ comes almost exclusively from
the diagonal term $\langle 0,1|A_2^0|0,1 \rangle$ and we
obtain the results listed in  Table \ref{SPLIT}.

\subsection{One-Phonon Manifold}

\subsubsection{Numerical Results}

Next we consider the manifold $J=1$ with $N=1$ phonon.
Again only $Y_l^m$ with $l=2$ contributes, so that we may write
\begin{eqnarray}
{\cal H}(1,1)_{\alpha \mu ; \alpha' \mu'} & \equiv & 
\langle \alpha \mu | {\cal H}(N=1,J=1) | \alpha' \mu' \rangle
\nonumber \\ &=& [2B + E_0 + \hbar \omega_\alpha ] \delta_{\mu , \mu'}
\delta_{\alpha , \alpha'} 
+ \langle 1 \alpha |A_2^{\mu-\mu'} ({\bf r})| 1 \alpha' \rangle
\langle 1 \mu |  Y_2^{\mu-\mu'}(\Omega) | 1 \mu' \rangle
\nonumber \\ && - \sum_{N' \not= 1} \sum_{\alpha'' =1}^{k_{N'}}
\sum_{\mu''} {1 \over E_{N', \alpha''} - E_{1,\alpha} }
\langle 1 \alpha | A_2^{\mu-\mu''}({\bf r}) | N' \alpha'' \rangle
\langle N' \alpha'' | A_2^{\mu''-\mu'}
({\bf r}) |1 \alpha' \rangle \nonumber \\
&& \ \ \times
\langle 1 \mu |  Y_2^{\mu-\mu''}(\Omega) | 1 \mu'' \rangle
\langle 1 \mu'' | Y_2^{\mu''-\mu'} (\Omega) | 1 \mu' \rangle \ .
\label{JEQ1EQ}
\end{eqnarray}
This is a $9 \times 9$ matrix, which gives the 9 ($J=1,N=1$) levels.
Since the first term gives rise to the removal of degeneracy expected
from group theoretical considerations, we did not include the
second term in our numerical evaluations.  This procedure was
sufficiently accurate to provide a useful check on the validity
of the more accurate numerical solutions for the three-component
wave functions of our set of mesh points.  In Table \ref{JN1}
these numerical results (``full mesh'') are given and are compared
to the results using perturbation theory, as in Eq. (\ref{JEQ1EQ}).
As can be seen, the two approaches yield quite compatible results.

\subsubsection{Qualitative Remarks}

Some additional comments on Eq. (\ref{JEQ1EQ}) are in order.
The first line of this equation gives the energy at first order
in perturbation theory.  At this order the wave function 
remains a product of the spatial ground-state
spatial wave function for
a $J=0$ molecule times a $J=1$ rotational wave function.  At
this level of approximation there is no dynamical coupling
between translation and rotation.  In second order perturbation
theory we see that admixtures of two phonon states which are
multiplied by different rotational states are introduced.
For example, consider the situation when the molecule
is in a uniaxial symmetry site and let
$|X\rangle$, $|Y \rangle$, and $|Z\rangle$ be the
$J=1$ states for which respectively $J_x$, $J_y$, and $J_z$
are zero.  If, for simplicity, we assume that the unperturbed
spatial wave function is spherically symmetric, then the state
which without perturbation was
\begin{eqnarray}
C_0|Z\rangle e^{-r^2/(4 \sigma^2)}
\end{eqnarray}
where $C_0$ is a normalization constant, is now
\begin{eqnarray}
C_1 |Z \rangle e^{-r^2/4 \sigma^2} 
+ C_2 |X\rangle zx e^{-r^2/(4 \sigma^2)}
+ C_2 |Y\rangle zy e^{-r^2/(4 \sigma^2)} \ ,
\label{C1ZEQ} \end{eqnarray}
where $C_1 \approx C_0$ and $C_2$ is small compared to $C_1$.
The point is that this formulation allows the molecule to change its
orientational state as it translates.  For H$_2$ in C$_{60}$ this
effect is small, however, in less symmetrical cavities,
as in nanotubes\cite{YHII}, this effect can become more important.

\subsubsection{Group Theoretical Analysis}

\begin{figure}
\centerline{\psfig{figure=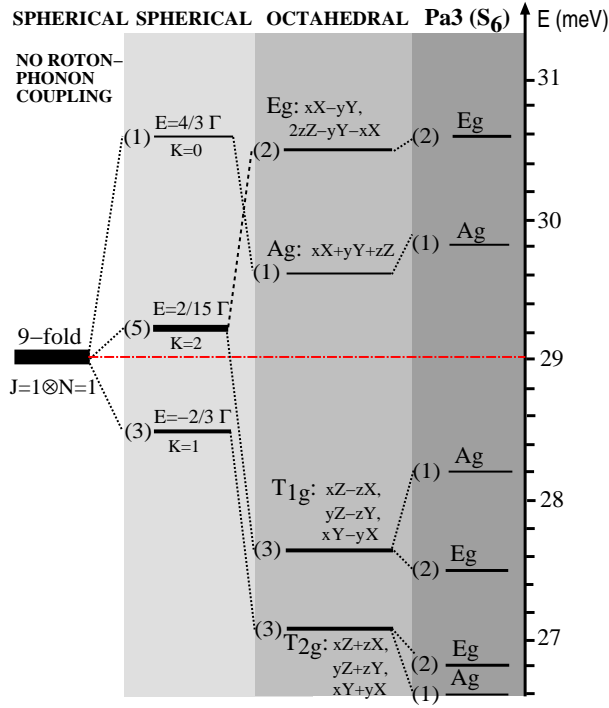,width=8cm}}

\vspace{0.2 in}
\caption{Removal of degeneracy as roton-phonon interactions are
introduced and the site symmetry is lowered. The degeneracy is indicated by
the number in parenthesis.  At the
far left is shown the completely degenerate level when spherical symmetry is
assumed and no roton-phonon coupling is present.  The next panel shows
the effect of allowing roton-phonon interactions but preserving overall
spherical symmetry.  Here K is the total (orbital plus orientational)
angular momentum. In the next panel spherical symmetry is lowered
to octahedral symmetry which is appropriate for H$_2$ in the octahedral
interstitial site in orientationally disordered C$_{60}$.
The far right panel
(and the energy scale) applies to the case of H$_2$ in orientationally ordered
C$_{60}$ in which case the site symmetry is S$_6$.}
\label{ORTHOENERGY} \end{figure}

In Fig. \ref{ORTHOENERGY} we show the influence of roton-phonon coupling and
local site symmetry on the energy spectrum of the one phonon-$(J=1)$
manifold.  At the far left we start from the case of highest
symmetry when the phonon and rotations separately have
complete rotational invariance and no phonon-roton coupling
is present.  In this case the manifold of 9 states
(3 one-phonon states $\otimes$ 3 $(J=1)$ states) is completely
degenerate.  When roton-phonon coupling is included (but
the environment is still spherically symmetric) we have
overall rotational invariance and the resulting eigenstates
are characterized by their total angular momentum $K$.  The
roton-phonon coupling causes states with different $K$ to have
different energy.  (The size of the splitting shown in the figure
is adjusted to agree with the center of gravity of the 
appropriate levels for cubic site symmetry.)

The two right-hand columns pertain to the situation when
a $(J=1)$ hydrogen molecule occupies the octahedral interstitial
site of C$_{60}$.  When the C$_{60}$ molecules are orientationally
disordered the interstitial site has O$_{\rm h}$ symmetry and
we consider that case first.
Use of the character tables for the O$_{\rm h}$ group indicates
that the original 9 dimensional reducible representation $\Gamma$
is decomposed into irreducible representation of the O$_{\rm h}$
group as
\begin{eqnarray}
\Gamma &=& T_{2 \rm g} \oplus T_{1 {\rm g}} \oplus E_{\rm g}
\oplus  A_{\rm g} \ .
\end{eqnarray}
and the basis functions associated with these irreducible representations
are given in Table \ref{SYM}.  As mentioned above, for temperatures below
about 260K, the C$_{60}$ molecules order into a structure of crystal symmetry
Pa${\bar 3}$\cite{pa1,pa2,pa3,pa4} in which case the formerly octahedral
interstitial has the lower S$_6$ symmetry.\cite{HAHN} Use of the relevant
character table shows that now
\begin{eqnarray}
\Gamma &=& 3A_{\rm g} \oplus 3 E_{\rm g} \oplus 3 E_{\rm g}^* \ ,
\end{eqnarray}
where $E_g$ is a complex one dimensional representation and $E_g^*$ is
its complex conjugate partner.  The basis functions associated with
these irreducible representations are given in Table \ref{SYM}.
The most important conclusion from this analysis is that the energy
eigenfunctions are {\it not} simply products of translational and
rotation wave functions, but instead are linear combinations of such
products.  This type of wavefunction reflects the fact that symmetry
operations act simultaneously on the position and the orientation of
a molecule.

\begin{figure}
\centerline{\psfig{figure=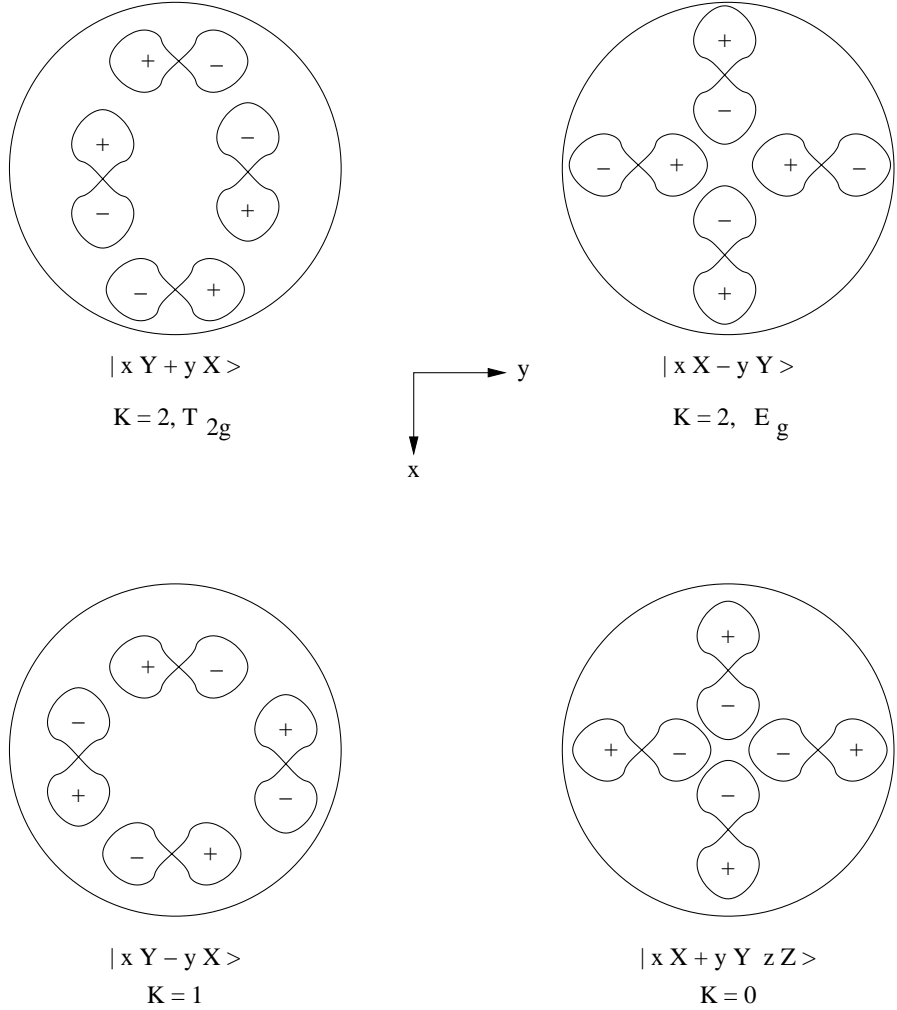,height=5.5 in}}

\vfill
\caption{Translation-rotation wavefunctions for
a $(J=1)$ H$_2$ molecule in an octahedral interstitial site
with one phonon.  Here the plane of the paper is the $x-y$
plane and for simplicity only the dependence in this plane is depicted.
Each figure eight represents an $|X\rangle$ or $|Y\rangle$
orientational wavefunction and the sign associated with each lobe
of this $p$-like function is indicated. Each
orientational wave function is multiplied by a translational
wavefunction $|x\rangle$, $|y\rangle$, or $|z\rangle$, where for
instance $|x \rangle \sim x \exp[- \case 1/4 (x/\sigma)^2]$.
The presence of a phonon in the $r_\alpha$ coordinate
thus causes the wave function to be an odd function of
$r_\alpha$, as one sees in the diagrams. As indicated 
in Fig. 2, the total angular momentum, $K$, which is the
sum of the angular momentum of the phonon and that of
rotation, is a good quantum number whose value is indicated.
Top, Left: a $K=2$, T$_{\rm 2g}$ function,
top right: a $K=2$, $E_{\rm g}$ function, bottom left: a $K=1$,
$T_{1 \rm g}$ function, and bottom right a $K=0$, A$_{\rm g}$ function.}
\label{TRFIG} 
\end{figure}

To emphasize this fact we give, in Fig. \ref{TRFIG}, a pictorial representations of the
translation-rotation wavefunctions.  This representation is to be
interpreted as follows.  We know that the rotational wave functions for
a free $J=1$ molecule can be taken to be analogs of $p_x$,
$p_y$ and $p_z$ functions and we will label these {\it rotational}
wavefunctions as $X$, $Y$, or $Z$.  For instance
$|X\rangle \sim \sin \theta \cos \phi$, $|Y\rangle \sim \sin \theta \sin \phi$,
and $|Z\rangle \sim \cos \theta$.  These wave functions have two lobes,
one positive the other negative aligned along the axis associated with their
state label.  When such a rotational function is multiplied by a one
phonon function in the $\alpha$ direction, ($|x\rangle$ denotes a
wave function for a single phonon excitation in the $x$-direction),
the total wave function will be an odd function of $\alpha$.  
Thus the wave function $| xX \rangle$ is an odd function of $x$ and
is therefore depicted by two $p_x$ functions, one at positive $x$
and another at $-x$ with the signs of the two lobes changed.
For simplicity  in the figures we
show only those functions which have appropriate dependence in
the plane of the paper, which is taken to be the $x$-$y$ plane.
At the upper left we show an $xy$-like function.  It has
two other t$_{2g}$ symmetry partners which are $xz$-like and
$yz$-like.  At the upper right we show one of the two E$_g$
functions which is $x^2-y^2$-like.  These five t$_{2g}$ and
E$_g$ functions comprise the manifold of total angular momentum
$K=2$ states.  Within spherical symmetry all five of these states
are degenerate in energy.  In the lower left of Fig. \ref{TRFIG}
we show the $z$-like function of $t_{1g}$ symmetry.  Its other
two partners are obtained by cyclically permuting $x$, $y$, and
$z$.  These three functions comprise the manifold of total
angular momentum $K=1$ states, which transform under rotation
as a vector.
Finally at the lower right we show the angular momentum $K=0$
state.  Thus in spherical symmetry, the nine $J=1$ single phonon
states give rise to three distinct energy manifolds which have
degeneracies 1, 3, and 5, corresponding respectively to total
angular momentum $K=0$, $K=1$, and $K=2$.

The simplest classical arguments do not reproduce the above
results.  For instance, one might argue that translation can occur
equivalently along either of three equivalent coordinate axes.
In each case, one can have the molecule oriented along the axis
of translational motion or perpendicular to that axis.  This
argument would suggest that the nine levels break into a three-fold
degenerate energy level in which the molecule is oriented
longitudinally and a six-fold level in which the molecule is
oriented transversely.  This discussion shows that it is
essential to treat the translation-rotation problem quantum\
mechanically to get the correct degeneracy.

\section{NEUTRON SPECTRUM}

In the experimental study of  Fitzgerald {\it et. al.}\cite{NIST},
neutron energy loss spectrum of H$_{2}$ trapped in C$_{60}$
was measured with an energy resolution of 0.3 meV. The spectrum
shows surprisingly rich features. However the origin of these
features  were not successfully identified in detail. Since
the observed neutron spectrum is a direct probe of the
intermolecular potential between H$_{2}$ molecules and C$_{60}$
host lattice, it is very important to see if available 
atom-atom potentials can give a  spectrum which is similar
to the experimental data.  A suitable analysis of the
high resolution inelastic neutron scattering data in 
Ref.\onlinecite{NIST} should, in principle, give 
a detailed information about the intermolecular
potential between H$_{2}$ molecules and the host lattice.

\begin{figure}
\centerline{\psfig{figure=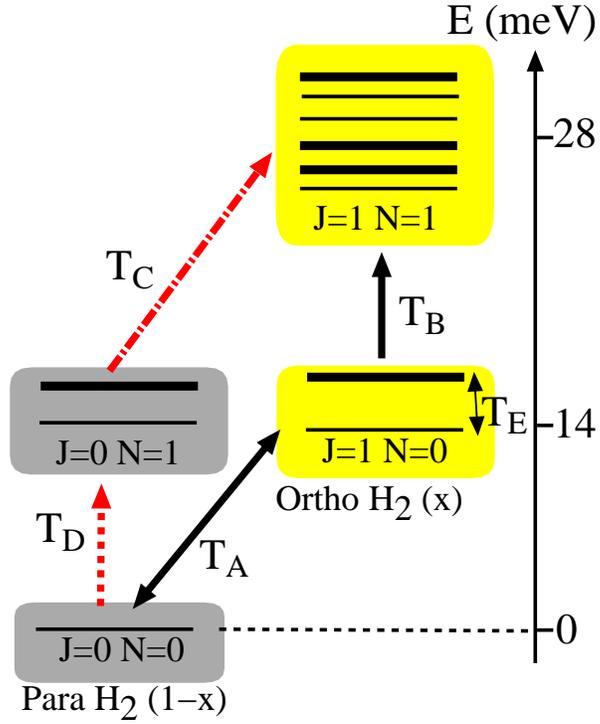,width=8cm}}
\vfill
\caption{A schematic representation of possible transitions between the
rotation-phonon energy levels that could be observed in a neutron
scattering experiment. At low temperature, only the $(J=0,N=0)$ and
$(J=1, N=0)$ states are populated and therefore the transition T$_{C}$ 
can not be observed at low temperature. The transition T$_{D}$ 
is  proportional to the coherent cross section of H$_{2}$ and
therefore very small. The transitions T$_{B}$ and T$_{A}$ have
comparable cross sections (see text for details).}
\label{TRANSITION}
\end{figure}

Figure~\ref{TRANSITION} illustrates several possible transitions,
involving both rotational and vibrational excitations, 
that could be observed
in a neutron scattering experiment for H$_{2}$ in solid C$_{60}$. 
In order to estimate
the intensities of these transitions and the corresponding neutron spectrum,
in Appendix \ref{NEUTRON} we derive the inelastic neutron cross 
section for trapped H$_2$ molecules in a powder sample at low temperature.
Below we discuss the contribution to the total neutron spectrum
from each of these transitions, labeled as $T_{A},...,T_{E}$, 
and then compare the calculated spectrum with experimental data
using various atom-atom potentials.

We start with the transitions involving phonon creation in
para hydrogen, as shown by $T_{D}$ in Fig.\ref{TRANSITION}.
Because of the spin-dependent interaction between the
proton and the neutron, processes in which a para molecule is
not converted to an ortho molecule are forbidden, or 
more correctly speaking, are proportional to the 
coherent cross section $b$, which is very small compared to
incoherent cross section $b'$. Hence, the transition $T_{D}$
will not have a noticeable contribution to the total cross
section. 

We next discuss the contribution to the total spectrum
from processes in which
either a $(J=0)$ molecule is converted to $(J=1)$ molecule
(para-ortho conversion, labeled as $T_{A}$ in Fig.~\ref{TRANSITION}) 
or a single phonon is created (as shown by $T_{B}$). Our 
calculations presented so far indicate that both processes
will give features around 14 meV in neutron spectrum.

In the Appendix~\ref{NEUTRON} we find that the cross section due to
para-ortho conversion,  (indicated by the subscript
$0 \rightarrow 1$ is given by
\begin{eqnarray}
{\partial^2 \sigma \over \partial \Omega \partial E} \Biggr)_{0 \rightarrow 1}
&=& \case 3/4 N {k' \over k} (1-x)  [b'j_1(\case 1/2 \kappa \rho)]^2
e^{-2W(\kappa)} \sum_m \delta [E_L - (E_c+E_m)] \ ,
\label{op}
\end{eqnarray}
where $N$ is the total number of H$_2$ molecules, ${\bf k}$ (${\bf k}'$) is
wave vector of the incident (scattered) neutron, $\kappav={\bf k}'-{\bf k}$,
$x$ is the fraction of H$_2$ molecules which are ortho (odd $J$)
molecules, $\rho$ is the separation between protons
in the H$_2$ molecule, $b'$ is the spin-dependent cross section in
the proton-neutron  pseudopotential,  $j_n$ is the $n$th order
spherical Bessel function, and $W(\kappa)$ is the Debye-Waller
factor which we take to be $\case 1/3 \kappa^2 \langle u^2 \rangle$.
Also, $E_L$ is the energy loss of the neutron, and $E_c+E_m$ is the
para-ortho conversion energy when the final state of the ortho has $J_z=m$.

Similarly, the cross section due to ortho-para conversion,
${\partial^2 \sigma \over \partial \Omega \partial E} \Biggr)_{1 \rightarrow 0}$
has the same expression as Eq. (\ref{op}) but now the factor $(1-x)$
is replaced by $x$. Hence 
the ratio of the total cross section for
ortho to para conversion to that of para to ortho conversion
is $(1-x)$ to $x$, where $x$ is the ortho concentration.
Normally the ratio of energy gain to energy loss cross sections
follows the Boltzmann factor.  Here, the populations are
set by $x$ rather than by the temperature.

We now discuss the cross section due to phonon
creation on a $(J=1)$ molecule (indicated here by the subscript
$1 \rightarrow 1$), which  is calculated in the Appendix~\ref{NEUTRON}.
These transitions are shown as $T_{B}$ in Fig.~\ref{TRANSITION}.
The result requires a knowledge of the
translation-rotation wave function of the H$_2$ molecule.  We find that
\begin{eqnarray}
{\partial^2 \sigma \over \partial \Omega \partial E}
\Biggl)_{1 \rightarrow 1} &=&
Nx {k' \over k} {b'}^2 \sum_{n=1}^4 S_{1 \rightarrow 1,n}^{(1)} \ ,
\end{eqnarray}
where the cross section $S_{1\rightarrow 1,n}^{(1)}$ are given in
Eqs. (\ref{I1EQ}), (\ref{I23EQ}), and (\ref{I4EQ}) of Appendix
\ref{NEUTRON}.

In Fig. \ref{PROBABILITIES} T$_{\rm B}$ represents the transitions
from the $(J=1,N=0)$ levels to the manifold of nine energy levels
of $(J=1, N=1)$.  Accordingly, we expect several transitions with
nonzero amplitude and thus rich features
in the total neutron cross section.

Also one may consider the cross section integrated over energy, which
is a  usefull quantity to indicate the relative strength of the
different transitions discussed above.  The ratio, $r_P$,
of the integrated neutron energy-loss cross section for phonon creation
(process T$_{\rm B}$ in Fig. ~ \ref{TRANSITION}) divided by that for
para-ortho conversion (process T$_{\rm A}$ in Fig.~\ref{TRANSITION})
was found in the Appendix~\ref{NEUTRON} to be
\begin{eqnarray}
r_P &=& \case 2/{27} \kappa^2 \langle u^2 \rangle  { x[ j_0(\case 1/2 \kappa \rho)^2
+2 j_2(\case 1/2 \kappa \rho)^2] \over (1-x) j_1(\case 1/2 \kappa \rho)^2 } \ .
\label{OPr}
\end{eqnarray}
This ratio is plotted as a function of $\kappa$ for $x=3/4$ in  
Fig.~\ref{O-P}.  Since this ratio is of order unity, the energy loss
spectrum will display features due to both phonons and
para-ortho conversion.

In appendix~\ref{NEUTRON} we also calculate the zero 
phonon ortho cross section 
for the transition shown as $T_{E}$ in Fig.~\ref{TRANSITION}. 
Since the $(J=1)$ levels are in thermal equilibrium, in this case the
ratio of the cross sections at energy gain to those of energy loss
does satisfy detailed balance. The ratio of the total cross
section (counting both energy loss and energy gain) for transitions
within the $(J=1)$ ground manifold to that due to para to ortho 
conversion was found at zero temperature to be
\begin{eqnarray}
r_{J=1} (T=0) &=& \frac{x}{1-x}  \frac{4 j_2(\case 1/2 \kappa \rho)^2}
{  15 j_1(\case 1/2 \kappa \rho)^2 } \ .
\label{rj}
\end{eqnarray}

Figure~\ref{O-P} 
shows that this ratio is quite small and therefore
experimental observation of this transition (i.e. $T_{E}$) would
be very difficult.

Figure~\ref{expvscal} shows the neutron energy loss spectrum 
and the calculated total spectrum using the same potential,
so called WS77 model,\cite{ATOM} 
used by FitzGerald {\it et. al}\cite{NIST}. 
Even though the calculated spectum is wider than the experimental
spectrum, it is still possible to make a one to one correspondence 
between calculation and experiment as is
shown by arrows in Fig.~\ref{expvscal}. The top curve 
in this figure shows what the spectrum looks like if the 
orientational part of the potential is scaled by about half.
The agreement between the data and calculations is somewhat better
after this arbitrary scaling, indicating that the potential
used is too anisotropic for the center of mass motion of 
H$_{2}$ molecule but at the same time it is too weak for the 
orientational part of H$_{2}$ (because it gives too small
a result for $\delta$, the splitting of the $(J=1)$ ground manifold).

We also tested other potentials commonly used in the literature
and these results are shown in Fig.~\ref{otherpot}.  
The top curve is from Novaco's 6-12 potential
which was developed to study hydrogen on graphite.\cite{novaco}
Clearly this potential gives too low phonon energies and too little
splitting of the $(J=1)$ levels for H$_{2}$ in solid C$_{60}$.
The other two curves in Fig.~\ref{otherpot} are 6-exp potentials
tabulated in Ref.\onlinecite{ATOM}. The spectrum from these
potentials does not agree with experiment either. 
We also searched the potential parameters $A$ and $B$ for 6-12
and $A$, $B$, and $C$ for 6-exp types of potentials. However we
were not able to improve the fit to experiment using these atom-atom
potentials. 
Hence, it seems that simple atom-atom potential does not describe the
details of the H$_{2}$-C$_{60}$ interaction well. It is an open
and important question to find a better potential can
reproduce the experimental spectrum better. It is also of important
to see how well the potentials obtained from 
density functional theory
within local density approximation will do. 

Finally, in addition to features observed near 14 meV,
Fitzgerald {\it et al.}\cite{NIST} also observed  a feature at about 28 meV 
in the energy gain time-of-flight spectrum. 
This energy is about twice that of either the
para-ortho conversion energy or the energy of the translational
phonon for an H$_2$ molecule in an octahedral interstitial site.
Clearly this feature represents the energy of two excitations,
but it was not clear whether these would be two phonons, one phonon
and one ortho-para transition or two ortho-para transitions.
In Fig. \ref{28FIG} we show the temperature dependence of the total
intensity of this feature.  This temperature dependence follows
a thermal activation with an energy of about 14 meV.  Thus the
initial state must consist of one thermally excited phonon
and the transition observed destroys one thermal phonon and
converts one ortho molecule (which occurs with temperature-independent
probability $x$) to a para molecule, thus giving the observed energy
of about 28 meV.

\section{Conclusions}

Our results indicate that the coupled phonon-roton problem
is a rich one.  For the light molecules of H$_2$ 
where the splitting of rotational levels is large compared to
most lattice vibrational modes, one is in a so-called weak-coupling
limit where the interactions between rotons and translational
phonons can be treated perturbatively.  Even at this level
one recovers an interesting structure.

Needless to say, we hope that the calculations in this paper
will motivate more detailed experiments at higher resolution
to elucidate the structure of the roton-phonon spectrum.

We may summarize our main conclusions as follows:

${\bullet} $  We have presented a systematic perturbative approach
to the calculation of the roton-phonon spectrum of hydrogen molecules
in confined geometry.  Our calculations agree with the group
theoretical analysis for the geometries considered here.

${\bullet} $  In a general way, the techniques of this paper
(use of the atom-atom potential combined with perturbation theory)
may prove useful to treat hydrogen molecules in other confined
geometries, in particular in or on nanotubes.  We are currently
analyzing this situation.

${\bullet} $  We give the first calculation of the expected energy loss
spectrum from hydrogen in C$_{60}$ in the energy range where
phonons and para-ortho conversion both are important. We find that
none  of the traditional 6-12 and 6-exp types of potentials give good
results for the detailed energy dependence of the observed phonon 
spectrum,  although the WS77
potential\cite{ATOM} we used was definitely the  most satisfactory.  
It is a theoretical challenge to determine a potential which fully 
reproduces the observed spectrum.

${\bullet} $ We identify the feature at 28 meV in the energy gain
spectrum as consisting of conversion of one ortho molecule
to a para molecule combined with annihilation of a single phonon.
This identification is uniquely indicated by the temperature dependence of
this feature.

\section{acknowledgement}
We acknowledge partial support from the Israel-US
Binational Science Foundation.

\newpage \begin{appendix}

\section{Atom-Atom Potential and $A_l^m$}
\label{YLM}

We model the interactions between the hydrogen molecule and the
surrounding cage of C$_{60}$ molecules using an atom-atom
potential.\cite{ATOM} Unless otherwise indicated, all the results 
reported in this paper are obtained from the same potential 
$ -A/r^{6}+B \exp(-Cr)$ that
is used in Ref.\onlinecite{NIST} (where 
$A=5.94$ eV \AA$^{6}$, $B= 678.2$ eV, and $C=3.67 $ \AA$^{-1}$).

We consider two cases depending on the orientational state
of C$_{60}$ molecules.  When the molecules in the surrounding cage are
orientationally disordered, we distribute the carbon centers uniformly
over the surface of a sphere. This case corresponds to the
octahedral symmetry discussed in the text.  
The resulting integration of the
atom-atom potential over a spherical surface is done analytically in
Ref.\onlinecite{NIST} and therefore is not given here.  
For the Pa${\bar 3}$ symmetry, the C$_{60}$ molecules are
rotated according to their Pa${\bar 3}$ settings and then the
total potential and $A_l^m$'s are calculated on a mesh points 
of a cube centered at the octahedral site. Below we 
derive a convenient way to obtain the $A_l^m$'s
from the atom-atom potential. 

We write the potential, $V_{\rm H-C}$
between a single H atom and a single C atom as
\begin{eqnarray}
V_{\rm H-C} &=& F( r )\ ,
\end{eqnarray}
where ${\bf r}$ is the displacement of the C atom relative to to
the H atom.  Thus the interaction, $V$(H$_2$) of a H$_2$ molecule with a
C atom can be written as
\begin{eqnarray}
V({\rm H}_2) &=& \sum_{\sigma=\pm 1}
F([r^2 + \case 1/4 \rho^2 + \sigma \rho {\bf r} \cdot \hat n]^{1/2} ) \ ,
\end{eqnarray}
where now ${\bf r}$ is the displacement of the C atom relative to the
center of the H$_2$ molecule whose atoms are at positions
$\pm \case 1/2 \rho \hat n$, where $\hat n$ is a unit vector specifying
the orientation of the molecular axis of the H$_2$ molecule.  Then
\begin{eqnarray}
A_{2l}^m &=& \sum_\sigma \sum_i  \int Y_{2l}^m( \hat n)^*
F([r_i^2 + \case 1/4 \rho^2 + \sigma \rho {\bf r}_i \cdot \hat n]^{1/2}) d \Omega \ ,
\label{A3} \end{eqnarray}
where $d \Omega$ indicates an integration over all orientations of $\hat n$
and the sum over $i$ is over all neighboring carbon atoms.  We use this to
get
\begin{eqnarray}
A_0^0 &\equiv& V_0 = 2\sqrt\pi \sum_i \int_{-1}^1 
F([r_i^2+\case 1/4 \rho^2
+ \rho r_i x]^{1/2}) dx \ .
\end{eqnarray}
To get the $A_l^m$'s for $l>0$ we write
\begin{eqnarray}
\sum_\sigma F([r_i^2 + \case 1/4 \rho^2 
+ \sigma \rho {\bf r}_i \cdot \hat n]^{1/2}) &=&
\sum_{L} B_{2L} (r_i) Y_{2L}^0(\theta_{r,n}) \ ,
\label{A5} \end{eqnarray}
where $\theta_{r,n}$ is the angle between the vectors ${\bf r}_i$ and $\hat n$.
We have that
\begin{eqnarray}
B_{2L}(r_i) &=& 2 \pi \sum_\sigma \int_0^\pi  F([r_i^2 + \case 1/4 \rho^2
+ \sigma \rho {\bf r}_i \cos \theta]^{1/2}) Y_{2L}^0(\theta) \sin \theta d \theta \ .
\label{A6} \end{eqnarray}
Now substitute Eq. (\ref{A5}) into Eq. (\ref{A3}) to get
\begin{eqnarray}
A_{2l}^m &=& \sum_i  \int Y_{2l}^m( \hat n)^*
\sum_L B_{2L}(r_i) Y_{2L}^0(\theta_{r,n}) d \Omega  \ .
\end{eqnarray}
Using the addition theorem for spherical harmonics\cite{ROSE} we have 
\begin{eqnarray}
A_{2l}^m &=& \sum_i  \int Y_{2l}^m( \hat n)^*
\sum_L B_{2L}(r_i) \sqrt{4 \pi/(2L+1)} \sum_\nu Y_{2L}^\nu(\hat n)
Y_{2L}^\nu(\hat {\bf r}_i)^*  d \Omega \nonumber \\
&=& \sum_i B_{2l}(r_i) \sqrt{4 \pi/(2l+1)} Y_{2l}^m (\hat {\bf r}_i)^* \ .
\end{eqnarray}
For $A_2^m$ we get
\begin{eqnarray}
A_2^m &=& 2 \pi \sum_i Y_2^m(\hat {\bf r}_i)^*
\int_{-1}^1 (3x^2-1) F([r_i^2+ \case 1/4 \rho^2 + r_i \rho x]^{1/2}) dx \ . 
\end{eqnarray}
For power-law functions $F$ (i.e. $ F \sim 1/r^{2n}$),
this integral can be done analytically.

\section{Spherical Cavity}
\label{CAVITY}

In this appendix we apply our formalism discussed in the text to a simple
toy model to utilize the main physics of quantum roton-phonon dynamics 
of a H$_{2}$ molecule confined in a 
spherically symmetric cavity for which the potential differs
perturbatively from harmonic.

\subsection{Orientationally Dependent Potential}

For a diatomic molecule for which the spherical part of the potential is
that of an isotropic spherical oscillator, we take
the orientationally dependent part of the potential to be
\begin{eqnarray}
U( {\bf r} , \Omega) &=& f(r) \left[ (\hat x \sin \theta \cos \phi
+ \hat y \sin \theta \sin \phi + \hat z \cos \theta )^2 - \case 1/3
\right] \ ,
\end{eqnarray}
where $\hat x \equiv x/r$, $\hat y \equiv y/r$, and $\hat z = z/r$.

This potential can be written in the canonical form of Eq. (\ref{AEQ}) with
\begin{eqnarray}
A_2^m &=& {8 \pi \over 15} f(r) Y_2^m (\hat r)^* \ .
\label{A2EQ} \end{eqnarray}

\subsection{Energy Shift of the $J=1$ Manifold}

To evaluate Eq. (\ref{CGEQ}) for the shift in the center of gravity of the
$J=1$ zero phonon levels, we need the wave function of the ground
state, namely
\begin{eqnarray}
\psi_0 ({\bf r}) &=& \alpha e^{- r^2/(4 \sigma^2)} \ .
\end{eqnarray}
where $\alpha = \sigma^{-3/2} (2 \pi)^{-3/4}$.  Also we can write the
two phonon excited states (for spherical symmetry) as
\begin{eqnarray}
\psi_{2,m}^{(L=2)} &=&
\beta (r/\sigma)^2 Y_2^m(\hat {\bf r}) e^{-r^2/(4 \sigma^2)} \ ,
\end{eqnarray}
for $m= -2, -1, 0, 1, 2$ and for angular momentum $L=2$ and where
$\beta = \sigma^{-3/2} \sqrt{2/15} (2 \pi)^{-1/4}$.
The sixth two-phonon state is an $s$-wave state, whose wave function we do
not need.  We will also need the $L=2$ (d-wave) four phonon states which are
\begin{eqnarray}
\psi_{4,m}^{(L=2)} &=&
\gamma [(r/\sigma)^4 - 7 (r/\sigma)^2]
Y_2^m (\hat {\bf r}) e^{-r^2/(4 \sigma^2)} \ ,
\end{eqnarray}
for $m= -2, -1, 0, 1, 2$ and for angular momentum $L=2$ and where
$\gamma = \sigma^{-3/2} \sqrt{1/105} (2 \pi)^{-1/4}$.

Now we assume the following polynomial fit for $f(r)$:
\begin{eqnarray}
f(r) = \gamma_2 (r/\sigma)^2 + \gamma_4 (r/\sigma)^4 + \gamma_6 (r/\sigma)^6
\ .
\label{POLY} \end{eqnarray}
We then find
\begin{eqnarray}
| \langle 0 | A_2^{m}({\bf r}) | \psi_{2,m}^{(L=2)} \rangle |
&=& 4 \sqrt{{\pi \over 15}} | \gamma_2 + 7 \gamma_4 + 63 \gamma_6 | 
\equiv 4 \sqrt{{\pi \over 15}} \Gamma 
\label{GAMEQ} \end{eqnarray}
and
\begin{eqnarray}
| \langle 0 | A_2^{m} ({\bf r}) | \psi_{4,m}^{(L=2)} \rangle |
&=& 4  \sqrt{{14 \pi \over 15}} | \gamma_4 + 18 \gamma_6 | 
\end{eqnarray}
Thus we have
\begin{eqnarray}
\Delta E &=& - {4 \over 15}
{(\gamma_2 + 7 \gamma_4 + 63 \gamma_6)^2 \over \hbar \omega}
- {28 \over 15} { (\gamma_4 + 18 \gamma_6)^2 \over \hbar \omega} \ .
\label{NUMERIC} \end{eqnarray}
From numerics we get
\begin{eqnarray}
A_2^0(z) = 0.07226 r^4 + 0.0348 r^6 \ .
\end{eqnarray}
For convenience we take $\sigma = 0.25 \AA$.  Then
$A_2^0$ is of the form of Eq. (\ref{POLY}) with
\begin{eqnarray}
\gamma_4 = 0.07226 \sigma^4 \ , \hspace{1 in}
\gamma_6 = 0.0348 \sigma^6 \ .
\end{eqnarray}
\begin{eqnarray}
\gamma_4 & = & 0.07226/256 = 0.282 \ {\rm meV} \\
\gamma_6 & = & 0.0348/4096 = 0.0084 \ {\rm meV} \ .
\end{eqnarray}
we evaluate the first and second terms on the right-hand side of Eq.
(\ref{NUMERIC}) to be 0.115 meV and 0.025 meV, respectively.
Thus we have
\begin{eqnarray}
\Delta E & = & 0.14 {\rm meV} \ .
\end{eqnarray}
Notice that $\Delta E$ is a very strong function of $\sigma$.
For instance, if you take $\sigma=0.26 \AA$, you get $\gamma_4=0.33$ meV
and $\gamma_6 =0.011$ meV in which case $\Delta E = 0.19$ meV.

\subsection{Effect of Translation-Rotation Coupling on the $J=1$ One Phonon Manifold}

In the absence of coupling between translations and rotations we characterize the
single-phonon states by phonon angular momentum $L_P$, so that
\begin{eqnarray}
|L_P=\pm 1 \rangle &=& \mp \left( {x \pm iy \over \sigma } \right) \psi_0({\bf r}) \ ,
\hspace{0.5 in} |L_P= 0 \rangle = (z/\sigma) \psi_0({\bf r}) \ .
\end{eqnarray}
We now wish to include the effect of the perturbation of the form of Eq. (\ref{AEQ})
when the coefficients are as in Eq. (\ref{A2EQ}), with $f(r)$ given by Eq. (\ref{POLY}).
We know that states are now characterized by the total angular momentum
${\bf K} = {\bf L_P} + {\bf J}$.  So the energy of the $K=2$ manifold is given by
\begin{eqnarray}
E(K=2) &=& \langle L_P=1, J_z=1 | U({\bf r}, \Omega) | L_P=1, J_z=1 \rangle \nonumber \\
&=& - {1 \over 15}
{ \int d {\bf r} \left| {x-iy \over \sigma } \right|^2
e^{- r^2/(2\sigma^2)} f(r) {3 z^2 - r^2 \over r^2}
\over \int d {\bf r}  \left| {x -iy \over \sigma } \right|^2
e^{- r^2/(2\sigma^2)} }
\langle J_z=1 |  (3J_z^2-2) | J_z=1 \rangle \nonumber \\
&=& {2 \int d {\bf r} r^2 f(r) e^{- r^2/(2\sigma^2)} \over
225 \sigma^2 \int d{\bf r} e^{- r^2/(2 \sigma^2)} } = \case 2/{15} \Gamma \ ,
\end{eqnarray}
where $ f(r)$ and $\Gamma$ are defined  in Eq. (\ref{POLY}) and
Eq. (\ref{GAMEQ}), respectively.

Similarly, one can evaluate the Hamiltonian in the manifold of states
$\phi_1 = |L_P=1, J_z=0\rangle$ and $\phi_2 = |L_P=0, J_z=1\rangle$.
The Hamiltonian matrix in this basis is found to be
\begin{eqnarray}
{\cal H} &=& \Gamma \left[ \begin{array} {c c }
- \case 4/{15} & \case 2/5 \\ \case 2/5 & - \case 4/{15} \\
\end{array} \right] \ .
\end{eqnarray}
This matrix has an eigenvalue $\case 2/{15}\Gamma$ which is associated 
with the $K=2$, $K_{z}=1$ state and 
the new eigenvalue for the $K=1$ manifold, $E(K=1) = - \case 2/3 \Gamma$.

Similarly, one can evaluate the Hamiltonian in the manifold of states
$\phi_1 = |L_P=1, J_z=-1\rangle$, $\phi_2 = |L_P=0, J_z=0\rangle$ and
$\phi_3 = |L_P=-1, J_z=1\rangle$.  The Hamiltonian matrix in this basis
is found to be
\begin{eqnarray}
{\cal H} &=& \Gamma \left[ \begin{array} {c c c }
\case 2/{15} & - \case 2/5 & \case 4/5 \\
- \case 2/5 & \case 8/{15} & - \case 2/5 \\
\case 4/5 & - \case 2/5 & \case 2/{15} \\
\end{array} \right] \ .
\end{eqnarray}
In this manifold we reproduce the eigenvalues for $K=2$ and $K=1$.  The new
eigenvalue is $E(K=0)=\case 4/3 \Gamma$.
\section{Neutron Scattering Cross Section}
\label{NEUTRON}

\subsection{GENERAL FORMULATION}

Following Elliott and Hartmann\cite{EH}, we write
\begin{eqnarray}
V({\bf r} - {\bf r}_n) &=& {2 \pi \hbar \over m_0} \delta ({\bf r}-{\bf r}_n)
\Biggl[ b + b' (\sigmav \cdot {\bf I} ) \Biggr] \ ,
\end{eqnarray}
where $m_0$ is the neutron mass, $\sigmav$ is the neutron spin, 
$\bf r$ is the coordinate of the proton, ${\bf r}_n$ the coordinate
of the neutron, $I$ is the proton spin, and $b$ and $b'$ are the
coherent and incoherent scattering lengths for this scattering.
Since $b$ is very small, we drop that term from now on.
The differential scattering cross section is
\begin{eqnarray}
{\partial^2 \sigma \over \partial \Omega \partial E} &=& {k' \over k}
\sum_{if} P_i \delta(E - E_i + E_f) |V|^2 \ ,
\end{eqnarray}
where $E=\hbar^2[(k')^2-k^2]/(2m_0)$ and the
sum is over all states of the system.  Here the potential is
\begin{eqnarray}
V = b' \sum_{j \alpha} e^{i \kappav \cdot {\bf R}_{j,\alpha}}
\sigmav \cdot I_{j,\alpha} \ ,
\end{eqnarray}
where $\kappav = {\bf k}'-{\bf k}$, $j$ labels molecules and
$\alpha=1,2$ the atoms within a molecule.  Performing the sum over
$\alpha$ we get
\begin{eqnarray}
V = b' \sum_j e^{i \kappav \cdot {\bf R}_j}
\Biggl( (\sigmav \cdot I_j) \cos( \case 1/2 \kappav \cdot \rhov )
+ i \sin(\case 1/2 \kappav \cdot \rhov) [\sigmav \cdot 
({\bf I}_{j1} - {\bf I}_{j2})] \Biggr) \ ,
\end{eqnarray}
where ${\bf I}_j = {\bf I}_{j1}+{\bf I}_{j2}$.  
The first term acts only on ortho molecules because for paras the
total spin is zero and the second
term causes transitions between ortho and para molecules.  So we write
the scattering cross section as the sum of three terms, the first of
which represents scattering from an ortho molecule and others
ortho-para conversion or the reverse.  Thus
\begin{eqnarray}
{\partial^2 \sigma \over \partial \Omega \partial E} &=& {k' \over k}
\Biggl[ Nx {\cal S}_{1 \rightarrow 1} + Nx {\cal S}_{1 \rightarrow 0}
+ N(1-x) {\cal S}_{0 \rightarrow 1} \Biggr] \ ,
\end{eqnarray}
where $N$ is the total number of molecules and $x$ is the ortho concentration.
Because there are no correlations
between nuclear spins each cross section is actually a sum over
cross sections for each molecule:
\begin{eqnarray}
{\cal S}_\beta &=& \sum_j {\cal S}_{\beta j} \ ,
\label{D6} \end{eqnarray}
where $j$ labels the molecule, $\beta$ is $0 \rightarrow 1$, etc., and
\begin{eqnarray}
{\cal S}_{0 \rightarrow 1 , j} & = &
\sum_{J_i=0,J_f=1} P_i \delta(E - E_i + E_f)
\Biggl|
\langle f | b' e^{i \kappav \cdot {\bf R}_j} \sin( \case 1/2 \kappav \cdot
\rhov) [\sigmav \cdot ({\bf I}_{j1} - {\bf I}_{j2})] | i \rangle_T
\Biggr|^2
\nonumber \\
{\cal S}_{1 \rightarrow 0 , j} & = &
\sum_{J_i=1,J_f=0} P_i \delta(E - E_i + E_f)
\Biggl|
\langle f | b' e^{i \kappav \cdot {\bf R}_j}
\sin( \case 1/2 \kappav \cdot \rhov) 
[\sigmav \cdot ({\bf I}_{j1} - {\bf I}_{j2})] | i \rangle_T \Biggr|^2
\nonumber \\ {\cal S}_{1 \rightarrow 1 , j} & = &
\sum_{J_i=1,J_f=1} P_i \delta(E - E_i + E_f)
\Biggl|
\langle f | b' e^{i \kappav \cdot {\bf R}_j}
\cos( \case 1/2 \kappav \cdot \rhov) [\sigmav \cdot {\bf I}] | i 
\rangle_T \Biggr|^2  \ ,
\end{eqnarray}
where the sums are over states for the fixed species (ortho or para) of
molecule as indicated and the subscript T indicates that the wavefunctions
include nuclear spin functions.

Now we perform the sum over the spin states of the neutron and proton to obtain the results
\begin{eqnarray}
{\cal S}_{0 \rightarrow 1 , j} & = & \case 3/4
(b')^2 \sum_{J_i=0,J_f=1} P_i \delta(E - E_i + E_f)
\Biggl|
\langle f | e^{i \kappav \cdot {\bf R}_j}
\sin( \case 1/2 \kappav \cdot \rhov) | i \rangle \Biggr|^2  \nonumber \\
{\cal S}_{1 \rightarrow 0 , j} & = & \case 1/4
(b')^2 \sum_{J_i=1,J_f=0} P_i \delta(E - E_i + E_f)
\Biggl|
\langle f | e^{i \kappav \cdot {\bf R}_j}
\sin( \case 1/2 \kappav \cdot \rhov) | i \rangle \Biggr|^2  \nonumber \\
{\cal S}_{1 \rightarrow 1 , j} & = & \case 1/2
(b')^2 \sum_{J_i=1,J_f=1} P_i \delta(E - E_i + E_f)
\Biggl|
\langle f | e^{i \kappav \cdot {\bf R}_j}
\cos( \case 1/2 \kappav \cdot \rhov) | i \rangle \Biggr|^2 \ ,
\end{eqnarray}
where now states $|f\rangle$ and $|i\rangle$ no longer include
nuclear spin wavefunctions.  We write
\begin{eqnarray}
e^{i \kappav \cdot {\bf R}_j} & =&
e^{i \kappav \cdot ( {\bf R}_j^{(0)} + {\bf u}_j)}
\approx e^{i \kappav \cdot {\bf R}_j^{(0)}} e^{- \case 1/2 (\kappav \cdot {\bf }u_j)^2}
\left[ 1 + i (\kappav \cdot {\bf u}_j) \right] \nonumber \\
&\approx& e^{-W} e^{i \kappav \cdot {\bf R}_j^{(0)}} \left[ 1 + i (\kappav \cdot {\bf u}_j)
\right] \ ,
\end{eqnarray}
where ${\bf R}_j^{(0)}$ is the equilibrium value of ${\bf R}_j$ 
and $W \approx \case 1/6 \kappa^2 \langle u^2 \rangle \equiv
\case 1/6 \kappa^2 \langle u_x^2 + u_y^2 + u_z^2 \rangle $.
Since spherical harmonics of degree higher than two do not affect the manifolds
of $J=0$ or $J=1$, we now use
\begin{eqnarray}
\cos ( \case 1/2 \kappav \cdot \rhov) &=&
j_0(\case 1/2 \kappa \rho) - 4 \pi j_2(\case 1/2 \kappa \rho)
\sum_\nu Y_2^n ( \hat \kappav )^* Y_2^n (\hat \rhov) 
\end{eqnarray}
and
\begin{eqnarray}
\sin ( \case 1/2 \kappav \cdot \rhov) = 4 \pi j_1(\case 1/2 \kappa \rho)
\sum_\nu Y_1^n ( \hat \kappav )^* Y_1^n (\hat \rhov) \ .
\end{eqnarray}

We expand in displacements to get
\begin{eqnarray}
{\cal S}_{0 \rightarrow 1 , j} & = & (4 \pi)^2 A
\sum_{J_i=0,J_f=1} P_i \delta(E - E_i + E_f)
\langle f | \left[ 1 + i \kappav \cdot {\bf u}_j \right]
\sum_\nu Y_1^\nu ( \hat \kappav )^* Y_1^\nu (\hat \rhov) | i \rangle
\nonumber \\ && \ \times
\langle i | \left[ 1 - i \kappav \cdot {\bf u}_j \right]
\sum_\mu Y_1^\mu ( \hat \kappav ) Y_1^\mu (\hat \rhov)^* | f \rangle \ ,
\end{eqnarray}
\begin{eqnarray}
{\cal S}_{1 \rightarrow 0 , j} & = & (4 \pi)^2 B
\sum_{J_i=1,J_f=0} P_i \delta(E - E_i + E_f)
\langle f | \left[ 1 + i \kappav \cdot {\bf u}_j \right] \nonumber \\ &&
\sum_\nu Y_1^\nu ( \hat \kappav )^* Y_1^\nu (\hat \rhov) | i \rangle
\langle i | \left[ 1 - i \kappav \cdot {\bf u}_j \right]
\sum_\mu Y_1^\mu ( \hat \kappav ) Y_1^\mu (\hat \rhov)^* | f \rangle \ ,
\end{eqnarray}
where $A=\case 3/4 e^{-2W}[b' j_1(\case 1/2 \kappa \rho)]^2$ and
$B= \case 1/4 e^{-2W} [b'j_1(\case 1/2 \kappa \rho) ]^2$ and
\begin{eqnarray}
{\cal S}_{1 \rightarrow 1 , j} & = &
\case 1/2 (b')^2 \sum_{J_i=1,J_f=1} P_i \delta(E - E_i + E_f) \nonumber \\ && \ \times
\langle f | \left[ 1 + i \kappav \cdot {\bf u}_j \right]
\left[ j_0(\case 1/2 \kappa \rho)
+ 4 \pi j_2(\case 1/2 \kappa \rho)
\sum_\nu Y_2^\nu ( \hat \kappav )^* Y_2^\nu (\hat \rhov) \right] | i \rangle
\nonumber \\ && \ \times
\langle i | \left[ 1 - i \kappav \cdot {\bf u}_j \right]
\left[ j_0(\case 1/2 \kappa \rho) + 4 \pi j_2(\case 1/2 \kappa \rho)
\sum_\mu Y_2^\mu ( \hat \kappav ) Y_2^\mu (\hat \rhov)^* \right] | f \rangle \ ,
\end{eqnarray}

Since the phonon energy is much larger than the orientational energy, we
may classify contributions according the number of phonons that are involved.
In the notation of Eq. (\ref{D6}) we write
\begin{eqnarray}
{\cal S}_\beta = {\cal S}_\beta^{(0)} + {\cal S}_\beta^{(1)} \ ,
\end{eqnarray}
where ${\cal S}_\beta^{(0)}$ corresponds to a zero-phonon process and
${\cal S}_\beta^{(1)}$ to a process in which one phonon is created or destroyed.
Thus
\begin{eqnarray}
{\cal S}_{0 \rightarrow 1}^{(0)} &=& (4 \pi)^2
A \sum_{J_i=0,J_f=1} P_i \delta(E - E_i + E_f) \sum_{\mu \nu}
\langle f | Y_1^\nu (\hat \rhov) | i \rangle
\langle i | Y_1^\mu (\hat \rhov)^* | f \rangle
Y_1^\nu (\hat \kappav)^* Y_1^\mu(\hat \kappav)
\end{eqnarray}
\begin{eqnarray}
{\cal S}_{0 \rightarrow 1}^{(1)} &=& (4 \pi)^2
A \sum_{J_i=0,J_f=1} P_i \delta(E - E_i + E_f) \nonumber \\ && \ \ \times
\sum_{\mu \nu \alpha \beta} \langle f | u_\alpha^* Y_1^\nu (\hat \rhov) | i \rangle
\langle i | u_\beta^* Y_1^\mu (\hat \rhov)^* | f \rangle
\kappa_\alpha \kappa_\beta Y_1^\nu (\hat \kappav)^* Y_1^\mu(\hat \kappav) \ .
\end{eqnarray}
Here and below we use spherical components of a vector ${\bf v}$:
$v_{\pm 1} = \mp (v_x\pm i v_y)/\sqrt 2$ and $v_0=v_z$.
Similarly
\begin{eqnarray}
{\cal S}_{1 \rightarrow 0}^{(0)} = (4 \pi)^2 B \sum_{J_i=1,J_f=0}
P_i \delta(E - E_i + E_f) \sum_{\mu \nu}
\langle f | Y_1^\nu (\hat \rhov) | i \rangle
\langle i | Y_1^\mu (\hat \rhov)^* | f \rangle
Y_1^\nu (\hat \kappav)^* Y_1^\mu(\hat \kappav) \ ,
\end{eqnarray}
\begin{eqnarray}
{\cal S}_{1 \rightarrow 0}^{(1)} &=& (4 \pi)^2 B \sum_{J_i=1,J_f=0} P_i
\delta(E - E_i + E_f)  \nonumber \\ && \ \ \times
\sum_{\mu \nu} \langle f | u_\alpha^* Y_1^\nu (\hat \rhov) | i \rangle
\langle i | u_\beta^* Y_1^\mu (\hat \rhov)^* | f \rangle
\kappa_\alpha \kappa_\beta Y_1^\nu (\hat \kappav)^* Y_1^\mu(\hat \kappav) \ ,
\end{eqnarray}
\begin{eqnarray}
S_{1 \rightarrow 1}^{(0)} &=& (4 \pi )^2 C \sum_{J_i=1,J_f=1}
P_i \delta(E - E_i + E_f) \sum_{\mu, \nu} \langle f | Y_2^\mu (\hat \rhov) | i \rangle
\langle i | Y_2^\nu ( \hat \rhov ) | f \rangle
Y_2^\mu (\hat \kappav)^* Y_2^\nu (\hat \kappav)^* \ ,
\label{11ZERO} \end{eqnarray}
where $C = \case 1/2 e^{-2W} [b' j_2(\case 1/2 \kappa \rho)]^2$ and
\begin{eqnarray}
&& {\cal S}_{1 \rightarrow 1}^{(1)} = 
D_0 \sum_{J_i=1,J_f=1} P_i \delta(E - E_i + E_f) \sum_{\alpha \beta}
\langle f | u_\alpha^* | i \rangle \langle i | u_\beta | f \rangle
\kappa_\alpha^* \kappa_\beta
\nonumber \\ && \ + 4 \pi D_1
\sum_{J_i=1,J_f=1} P_i \delta(E - E_i + E_f) \sum_{\mu \alpha \beta}
\langle f | u_\alpha^* Y_2^\mu (\hat \rhov) | i \rangle \langle i | u_\beta^* | f \rangle
\kappa_\alpha \kappa_\beta Y_2^\mu (\hat \kappav)^*
\nonumber \\ && \ + 4 \pi D_1
\sum_{J_i=1,J_f=1} P_i \delta(E - E_i + E_f) \sum_{\mu \alpha \beta}
\langle f | u_\alpha^* | i \rangle
\langle i | u_\beta^* Y_2^\mu(\hat \rhov) | f \rangle
\kappa_\alpha \kappa_\beta Y_2^\mu (\hat \kappav)^*
\nonumber \\ && \ + (4 \pi)^2 D_2
\sum_{J_i=1,J_f=1} P_i \delta(E - E_i + E_f) \sum_{\mu \nu \alpha \beta}
\langle f | u_\alpha^* Y_2^\nu(\hat \rhov)^* | i \rangle 
\langle i | u_\beta^* Y_2^\mu(\hat \rhov) | f \rangle
\kappa_\alpha \kappa_\beta Y_2^\mu (\hat \kappav)^* Y_2^\nu (\hat \kappav) \nonumber \\
&=& \sum_{n=1}^4 S_{1 \rightarrow 1;n}^{(1)} \ ,
\label{11ONE}\end{eqnarray}
where $D_n = \case 1/2 (b')^2 e^{-2W} j_0(\case 1/2 \kappa \rho)^{2-n} j_2(\case 1/2 \kappa
\rho)^n$ and $S_{1 \rightarrow 1;n}^{(1)}$ is the contribution to the one-phonon
ortho cross section from the $n$th term in the first equality.

Note the existence of terms in which a phonon and a roton are created, the system 
evolves and finally a phonon is destroyed.  This type of process can only
occur when the system supports roton-phonon interactions.  All the terms
${\cal S}_{1 \rightarrow 1;n}^{(1)}$ correspond approximately to the phonon energy.

\subsection{POWDER AVERAGE AT LOW TEMPERATURE}

Here we restrict attention to the energy loss spectrum at low temperature when
there are no thermal phonons present.  Also, we now take the powder average.
This corresponds to actual experimental situation
in Ref. \onlinecite{NIST}, but would also be a reasonable approximation to take
account of the differently oriented symmetry axes of the 
various octahedral interstitial sites.
Below we calculate the cross sections for the following processes;
(a) energy loss by conversion, (b) energy gain by conversion,
(c) single phonon energy loss, and finally (d) zero phonon
transition from $(J=1,M=0)$ to $(J=1,M=\pm1)$. 

\subsubsection{Energy Loss by Conversion}

We have
\begin{eqnarray}
\langle {\cal S}_{0 \rightarrow 1}^{(0)} \rangle  & = & (4 \pi) A
\sum_{J_i=0,J_f=1} P_i \delta(E - E_i + E_f) \sum_\mu
\langle f | Y_1^\mu (\hat \rhov) | i \rangle
\langle i | Y_1^\mu (\hat \rhov)^* | f \rangle \ , 
\end{eqnarray}
where $\langle \ \ \rangle$ indicates a powder average.
The initial state is the $J=0,J_z=0$ zero-phonon state, which we write as
\begin{eqnarray}
\psi_i &=& \sum_{\bf r} c_i({\bf r}) |{\bf r}; J=0; J_z=0 \rangle \ ,
\end{eqnarray}
where $c_i({\bf r})$ is the amplitude of the wave function at the mesh point ${\bf r}$.
The final state is a $J=1$ zero-phonon state, which we similarly write as
\begin{eqnarray}
\psi_{f,m} &=& \sum_{{\bf r},M} c_{f,m}({\bf r},M) |{\bf r}; J=1; J_z=M \rangle
\end{eqnarray}
and whose energy relative to the $J=0$ state is 
\begin{eqnarray}
E_{f,m} = E_c + E_m \ .
\end{eqnarray}
If $E_L=-E$ is the energy loss, we may write
\begin{eqnarray}
\langle {\cal S}_{0 \rightarrow 1}^{(0)}(E_L)  \rangle  & = & A \sum_m \delta [E_L - (E_c + E_m )]
\sum_\mu \Biggl| \sum_{\bf r} c_{f,m}({\bf r},\mu)^* c_i({\bf r}) \Biggr|^2 \ .
\end{eqnarray}
To a good approximation the zero-phonon wave functions for $J=1$ can be chosen to be
composed of a single value of $J_z$.  
Thus we may label the wave functions so that $\mu=m$:
\begin{eqnarray}
\langle {\cal S}_{0 \rightarrow 1}^{(0)}(E_L)  \rangle  & = & A \sum_m \delta [E_L - (E_c + E_m )]
\Biggl| \sum_{\bf r} c_{f,m}({\bf r},m)^* c_i({\bf r}) \Biggr|^2 \ .
\end{eqnarray}
The corresponding integrated intensity is
\begin{eqnarray}
I^{(0)} & \equiv & \int dE_L \langle {\cal S}_{0 \rightarrow 1}^{(0)} \rangle
= A \sum_m \Biggl| \sum_{\bf r} c_{f,m}({\bf r},m)^* c_i({\bf r}) \Biggr|^2 \ .
\end{eqnarray}
The inner product of the $J=0$ wave function and the spatial part of the 
$J=1$ zero-phonon states is essentially unity.  So
\begin{eqnarray}
I^{(0)} & = & 3A \ .
\end{eqnarray}

\subsubsection{Energy Gain by Conversion}

Here we give a similar analysis of the energy-gain spectrum at low
temperature due to ortho-para conversion.  The derivation is
similar to that for para-ortho conversion so we only
quote the results:
\begin{eqnarray}
\langle {\cal S}_{1 \rightarrow 0}^{(0)}(E_L)  \rangle  & = &
B \sum_m P_m \delta [E - (E_c - E_m )]
\Biggl| \sum_{\bf r} c_{i,m}({\bf r},m)^* c_f({\bf r}) \Biggr|^2 \ ,
\end{eqnarray}
where $P_m$ is the probability that the $J=1, J_z = m$ state is occupied
and the role of initial and final states is interchanged from
the para to ortho processes.  The corresponding integrated intensity is
\begin{eqnarray}
I^{(0)} & \equiv & \int dE \langle {\cal S}_{1 \rightarrow 0}^{(0)} \rangle
= B \sum_m  P_m \Biggl|  \sum_{\bf r} c_{i,m}({\bf r},m)^* c_f({\bf r})
\Biggr|^2 \nonumber \\
&=& B \ .
\end{eqnarray}
Since $B=3A$ we see that ratio of the total cross section for
ortho to para conversion to that of para to ortho conversion
is $(1-x)$ to $x$, where $x$ is the ortho concentration.
Normally the ratio of energy gain to energy loss cross sections
follows the Boltzmann factor.  Here, the populations are
set by $x$ rather than by the temperature.

\subsubsection{Single Phonon Energy Loss}

We have the powder average of $S_{1 \rightarrow 1;1}^{(1)}$ of
Eq. (\ref{11ONE}) as
\begin{eqnarray}
\langle S_{1 \rightarrow 1;1}^{(1)} \rangle &=&
\case 1/3 \kappa^2 D_0 \sum_{J_i=1,J_f=1} P_i \delta(E - E_i + E_f) \sum_L
| \langle f | u_L | i \rangle |^2 
\end{eqnarray}
and the corresponding integrated intensity is
\begin{eqnarray}
I^{(1)}_1 &\equiv& \int dE_L \langle S_{1 \rightarrow 1;1}^{(1)} (E_L)\rangle
= \case 1/3 \kappa^2 \langle u^2 \rangle D_0 \ . 
\end{eqnarray}
In terms of the amplitudes of the wave function on the mesh points,
the above result is
\begin{eqnarray}
\langle S^{(1)}_{1 \rightarrow 1;1} \rangle
&=& \case 1/3 \kappa^2 D_0 \sum_{i,f} P_i \delta(E_L-E_f+E_i)
\sum_{L} \left| \sum_{M, {\bf r}} c_i(M,{\bf r}) c_f(M,{\bf r}) r_L \right|^2 \ .
\label{I1EQ} \end{eqnarray}

Also we obtain the powder average of $S_{1 \rightarrow 1;2}^{(1)}$ and
$S_{1 \rightarrow 1;3}^{(1)}$ of Eq. (\ref{11ONE}) as
\begin{eqnarray}
\langle S_{1 \rightarrow 1;2}^{(1)}(E_L) \rangle &=& 
\langle S_{1 \rightarrow 1;3}^{(1)}(E_L) \rangle^* =
- {2 \over 5} \sqrt{ {8 \pi \over 15}} D_1 \kappa^2 \sum_{i,f} P_i \delta(E_L-E_f+E_i) \nonumber \\
&& \ \times \sum_{M,M'} C(112;M,M') (-1)^{M+M'} \langle f| u_{-M} T_2^{M+M'}({\bf J}) | i \rangle
\langle i| u_{-M'}| f \rangle \ ,
\label{I23EQ} \end{eqnarray}
where the $T_2^M({\bf J})$ are the operator equivalents of the spherical harmonics:
\begin{eqnarray}
T_2^{\pm 2}({\bf J}) &=& \sqrt{{15 \over 32 \pi}} J_\pm^2 \ , \ \ \
T_2^{\pm 1}({\bf J}) = \mp \sqrt{{15 \over 32 \pi}} (J_z J_\pm + J_\pm J_z) \ , \ \ \
T_2^0({\bf J}) = \sqrt{{ 5 \over 16 \pi}} (3J_z^2-2) \ .
\end{eqnarray}
Here
\begin{eqnarray}
\langle i| u_{-M'}| f \rangle &=& \sum_{\mu,{\bf r}} c_i(\mu,{\bf r})^* c_f(\mu,{\bf r})        r_{-M'} \ .
\end{eqnarray}
and
\begin{eqnarray}
\langle f| u_{-M} T_2^{M+M'}({\bf J}) | i \rangle &=& {5 \over \sqrt{8 \pi}}
\sum_{\mu , {\bf r}} c_i(\mu, {\bf r}) c_f(\mu+M+M',{\bf r})^*
r_{-M} C(121;\mu,M+M') \ ,
\end{eqnarray}
where we used
\begin{eqnarray}
\langle J=1;J_z=M | T_2^L(\rhov) | J=1; J_z=M' \rangle &=&
{5 \over \sqrt{8 \pi}} \delta_{M,L+M'} C(121;M',L) \ .
\end{eqnarray}

We have the contributions to the integrated intensity
\begin{eqnarray}
I^{(1)}_2 &= & {I^{(1)}_3}^* = \int dE_L \langle S_{1 \rightarrow 1;2}^{(1)} (E_L)
\rangle \nonumber \\ &=&
- {2 \over 5} \sqrt{{8 \pi \over 15}} D_1 \kappa^2 \sum_{if}
P_i \sum_{M,M'} C(112;M,M') (-1)^{M+M'} \langle i |
u_{-M'} |f \rangle \langle f | u_{-M} T_2^{M+M'} ({\bf J}) |i \rangle \ .
\end{eqnarray}
Here the sum over final states should be restricted to single phonon
states.  Higher energy states make only a small contribution to
this sum.  So we make the closure approximation that the sum over
$|f\rangle$ extends over all states, in which case
\begin{eqnarray}
I^{(1)}_2 &=& - {2 \over 5} \sqrt{{8 \pi \over 15}} D_1 \kappa^2 \sum_{i}
P_i \sum_{M,M'} C(112;M,M') (-1)^M \langle i |
u_{M'}^* u_{-M} T_2^{M+M'} ({\bf J}) |i \rangle \ .
\label{C41}
\end{eqnarray}
As illustrated by Eq. (\ref{C1ZEQ}), the initial state $|i\rangle$
is dominantly comprised of a single value of $J_z$.  Thus
in Eq.~\ref{C41} $M+M'=0$ dominates.  
In addition, the system is nearly isotropic.
Then $\sum_M C(112;M,-M)(-1)^M |u_{-M}|^2=0$. So, to a good approximation,
\begin{eqnarray}
I^{(1)}_2 + I^{(1)}_3=0 \ .
\end{eqnarray}
We have made several approximations, but our result for the total
integrated intensity will not be much affected by these approximations.

Similarly, we get
\begin{eqnarray}
&\langle &S_{1 \rightarrow 1;4}^{(1)} (E_L) \rangle
= {16 \pi \over 75} \kappa^2 D_2 \sum_{i, f, \mu , M}
P_i \delta(E_L-E_f + E_i) \Biggl| \langle f| u_M T_2^\mu({\bf J}) |i \rangle \Biggr|^2
\nonumber \\
&& - {32 \pi \over 25 \sqrt{21}} D_2 \kappa^2 \sum_{i, f, \nu, M, M'} P_i
\delta(E_L-E_f+E_i) C(112;M,M') C(222;M+M',\nu) (-1)^{M'}
\nonumber \\ && \ \times 
\langle f| u_{-M} T_2^{-\nu}({\bf J}) |i \rangle \langle f | u_{M'} T_2^{-M-M'-\nu}({\bf J}) | i \rangle^* \ .
\label{I4EQ} \end{eqnarray}

We now evaluate the integrated intensity
\begin{eqnarray}
I^{(1)}_4 & \equiv & \int dE_L \langle S_{1 \rightarrow 1;4}^{(1)}
(E_L) \rangle \equiv t_1 + t_2 \ ,
\end{eqnarray}
where
\begin{eqnarray}
t_1 &=& {16 \pi \over 75} \kappa^2 D_2 \sum_{i,f,\mu,M}
P_i \langle i | u_{-M} T_2^{-\mu} ({\bf J}) |f\rangle 
\langle f| u_M T_2^\mu({\bf J}) |i \rangle (-1)^{M+\mu} \ .
\end{eqnarray}
Making the closure approximation this is
\begin{eqnarray}
t_1 &=& {16 \pi \over 75} \kappa^2 D_2 \sum_{i,\mu,M} P_i
\langle i | u_{-M} u_M T_2^{-\mu}({\bf J}) T_2^\mu({\bf J})
|i \rangle (-1)^{\mu+M} \nonumber \\ &=&
\case 2/3 \kappa^2 D_2 \sum_i P_i \langle i
| u_{-M} u_M |i \rangle (-1)^M = \case 2/3 \kappa^2 D_2
\langle u^2 \rangle \ .
\end{eqnarray}
Similarly
\begin{eqnarray}
t_2 &=& - {32 \pi \over 25 \sqrt{21}} D_2 \kappa^2 \sum_{i,M,M'\nu}
P_i C(112;M,M') C(222;M+M',\nu) \nonumber \\ && \ \times (-1)^{M+\nu}
\langle i | u_M^* T_2^{M+M'+\nu}({\bf J}) u_{-M} T_2^{-\nu}({\bf J})
|i \rangle \ .
\end{eqnarray}
Again, we treat $|i\rangle$ as having a single value of $J_z$,
so that $M+M'=0$.  Also we again assume spatial isotropy, so
that $\sum_M C(112;M,-M)(-1)^M |u_{-M}|^2=0$. Then $t_2=0$.

\subsubsection{Zero Phonon Ortho Cross Section}

Finally we consider the zero phonon contribution to the
ortho to ortho cross section.  Taking the powder average,
we have
\begin{eqnarray}
\langle S_{1 \rightarrow 1}^{(0)} \rangle &=& 4 \pi C \sum_{J_i=1}
\sum_{J_f=1} P_i \delta (E-E_i + E_f) \sum_\mu
\langle f| Y_2^\mu ( \hat \rhov  ) | i \rangle
\langle i| Y_2^{-\mu} ( \hat \rhov  ) | f \rangle
(-1)^\mu \ .\end{eqnarray}
If $\delta$ is the energy of the $J_z=\pm 1$ levels relative
to the $J_z=0$ level and if the Boltzmann probability of
the state $|m>$ is $P_m$, then we may write
\begin{eqnarray}
\langle S_{1 \rightarrow 1}^{(0)} \rangle &=& 4 \pi C P_0 \delta (E+\delta)
\sum_{m= \pm 1} \Biggl[ | \langle m | Y_2^1(\hat \rhov) | 0 \rangle |^2
+ | \langle m | Y_2^{-1}(\hat \rhov) | 0 \rangle |^2 \Biggr]
\nonumber \\ && \ +
4 \pi C \delta(E-\delta) \sum_{m= \pm 1} P_m \Biggl[
| \langle m | Y_2^1(\hat \rhov) | 0 \rangle |^2
+ | \langle m | Y_2^{-1}(\hat \rhov) | 0 \rangle |^2 \Biggr] \ ,
\end{eqnarray}
which gives
\begin{eqnarray}
\langle S_{1 \rightarrow 1}^{(0)} \rangle &=&
\case 6/5 C P_0 \delta (E+\delta) + \case 6/5 C P_1 \delta (E-\delta) \ .
\end{eqnarray}
The ratio  of cross sections at energy gain to those of energy loss
does satisfy detailed balance because within the species
$(J=1)$ we do maintain thermal equilibrium.

 \subsection{INTENSITY RATIOS}

We develop an expression for the ratio, $r_P$, defined to be the integrated
intensity due to phonon creation divided by that due to para-ortho
conversion.  Using the results for $I^{(n)}$ obtained above we have
\begin{eqnarray}
r_P &=& { x \int dE_L \langle S_{1 \rightarrow 1}^{(1)} (E_L)\rangle 
\over (1-x) \int dE_L \langle S_{0 \rightarrow 1}^{(0)} (E_L) \rangle }
= {x[I^{(1)}_1+I^{(1)}_4] \over (1-x) I^{(0)} } \nonumber  \\
&=& { \case 1/3 \kappa^2 \langle u^2 \rangle D_0 + \case 2/3 \kappa^2 \langle u^2 \rangle
D_2 \over 3A} \left( {x \over 1-x} \right)  \nonumber \\ &=&
\case 2/{27} \kappa^2 \langle u^2 \rangle { j_0(\case 1/2 \kappa \rho)^2
+2 j_2(\case 1/2 \kappa \rho)^2 \over j_1(\case 1/2 \kappa \rho)^2 }
\left( {x \over 1-x} \right) \ .
\end{eqnarray}
The ratio  $r_C$, defined to be the ortho to para conversion
cross section in energy gain to that in energy loss due
to para to ortho conversion, is given by
\begin{eqnarray}
r_C &=& {x \over 1-x} \ .
\end{eqnarray}

Finally, we have  $r_{J=1}$, defined to be the
total cross section (counting both energy
loss and energy gain) for transitions within the
$J=1$ ground manifold divided by that due to para to
ortho conversion, as
\begin{eqnarray}
r_{J=1} = \biggl( \frac{x}{1-x} \biggr )
 { \case 6/5 C (P_0+P_1)\over 3A } = (P_0+P_1) r_{J=1} (T=0) \ ,
\end{eqnarray} 
where
\begin{eqnarray}
r_{J=1} (T=0) = \biggl( \frac{x}{1-x} \biggr )
{4 j_2(\case 1/2 \kappa \rho)^2 \over 15 j_1(\case 1/2 \kappa \rho)^2 } \ .
\end{eqnarray} 

\end{appendix}

\begin{table}
\caption{Effect of a spherical symmetric perturbation on $U_3$ states.
The $U_3$ states are characterized by $N$, the total number of harmonic phonons.
Wave functions in a spherical potential are characterized by angular
momentum $K$.  Here we give the effect of the perturbation
$\Delta (r/\sigma)^4$ , where $\langle r^2 \rangle = 3 \sigma^2$
for the isotropic harmonic oscillator in three spatial dimensions. 
In the last column we give the shift in the average energy of the
multiplet of states of a given value of $N$.}

\vspace{0.2 in} \noindent
\begin{tabular} { | c c c c | }
$N$ & $N,K$ & Energy & Avg $E$ \\
\hline \hline
0 & (0,0) & 15 $\Delta$ & 15 $\Delta$ \\
\hline \hline
1 & (1,1) & 35 $\Delta$ & 35 $\Delta$ \\ 
\hline \hline
2 & (2,0) & 75 $\Delta$ & 65 $\Delta$ \\
  & (2,2) & 63 $\Delta$ & \\
\hline \hline
3 & (3,1) & 119 $\Delta$ & 105 $\Delta$ \\
  & (3,3) &  99 $\Delta$ & \\
\end{tabular}
\label{AHTAB} \end{table}

\newpage 
\begin{table}
\caption{Energy level systematics for a $(J=0)$ molecule in an octahedral
interstitial site of C$_{60}$.  Here we show the removal
of degeneracy from a manifold of initial symmetry
${\cal I}$ to manifolds of final symmetry ${\cal F}$ due to a perturbation
$V$, as calculated in lowest-order perturbation theory.  Here $N$ is the
total number of phonons, $K$ is the angular momentum, and the other
group theoretical labels are as in Fig. \ref{PARAENERGY}.  We give a typical
eigenfunction $\psi$ to illustrate the symmetry.  Here $d$ is the
degeneracy (Deg) of the manifold and $\sigma^2= \langle x^2 \rangle =
\langle y^2 \rangle = \langle z^2 \rangle$.  Here the coordinate axes
coincide with the cubic [100] directions.}
\vspace{0.2 in} \noindent
\begin{tabular} { | c | c || c || c | c || c | c |}
${\cal I}$ & $\psi$ &  $V$ & $\psi$ & ${\cal F}$ & Deg & Energy \\
\hline \hline
$N=2,K=2$ & $r^2Y_2^M(\Omega)$ & $\kappa (x^4+y^4+z^4- \case 3/5 r^4)$ & 
$(x^2-y^2)$ & E$_{\rm g}$ & 2 & $\case {36}/5 \kappa \sigma^4$ \\
& & & $xy$ & T$_{2\rm g}$ & 3 & $- \case {24}/5 \kappa \sigma^4$ \\
\hline
$N=3,K=3$ & $r^3 Y_3^M(\Omega)$ & $\kappa (x^4+y^4+z^4- \case 3/5 r^4)$ & 
$(x^3- \case 3/5 xr^2)$ & T$_{1 \rm u}$ & 3 & $\case {36}/5 \kappa \sigma^4$ \\
& & & $x(y^2-z^2)$ & T$_{2 \rm u}$ & 3 & $- \case {12}/5 \kappa \sigma^4$ \\
& & & $xyz$ & A$_{\rm u}$ & 1  & $- \case {72}/5 \kappa \sigma^4$ \\
\hline \hline
$N=1$,$K=1$, T$_{1\rm u}$ & $rY_1^M(\Omega)$ & $\lambda (xy+yz+zx)$ &
& E$_{\rm u}$ & 2 & $- \lambda \sigma^2$ \\ 
& & & & A$_{\rm u}$ & 1 & $2 \lambda \sigma^2$ \\ 
\hline
$N=2$,$K=2$, T$_{2\rm g}$ & $xy$ & $\lambda (xy+yz+zx)$ & 
& E$_{\rm g}$ & 2 & $- \lambda \sigma^2$ \\ 
& & & & A$_{\rm g}$ & 1 & $2 \lambda \sigma^2$ \\ 
\hline
\hline
$N=3$,$K=3$, T$_{2 \rm u}$ & $x(y^2-z^2)$ & $\lambda (xy+yz+zx)$ &
& E$_{\rm u}$ & 2 & $\case 3/2 \lambda \sigma^2$ \\ 
& & & & $A_{\rm u}$ & 1 & $-3 \lambda \sigma^2$ \\ 
\hline
$N=3$,$K=3$, T$_{1 \rm u}$ & $(x^3-\case 3/5 xr^2)$ & $\lambda (xy+yz+zx)$ &
& E$_{\rm u}$ & 2 & $\case 3/{10} \lambda \sigma^2$ \\ 
& & & & A$_{\rm u}$ & 1 & $- \case 3/5 \lambda \sigma^2$ \\ 
\hline
$N=3$,$K=1$, T$_{1 \rm u}$ & $x(r^2- \case 1/3 \sigma^2)$
& $\lambda (xy+yz+zx)$ &
& E$_{\rm u}$ & 2 & $3 \lambda \sigma^2$\\ 
& & & & A$_{\rm u}$ & 1 & $-6 \lambda \sigma^2$  \\ 
\end{tabular}
\label{DEGENERACY}
\end{table}

\vspace{0.2 in}
\begin{table}
\caption{Phonon levels for a $(J=0)$ H$_{2}$ molecule in an octahedral site
for orientationally ordered and disordered C$_{20}$ respectively. 
Energies (meV) are with respect to the ground state energy E$_{0,1}$.
The symmetry of each manifold of degenerate levels can be read from 
Fig. 1.}

\vspace{0.2 in}
\begin{tabular}{crcc}  
\multicolumn{2}{c}{E$_{N,\alpha}$ } & Octahedral 
& Pa${\bar 3}$ (S$_{6}$) \\ \cline{1-2}
N & $\alpha$ & & \\ \hline
1 & 1 & 14.38 & 13.16 \\ 
1 & 2,3 & 14.38 & 14.47 \\ \hline 
2 & 1 & 28.26 & 26.69  \\ 
2 & 2,3 & 28.26 & 27.49 \\ 
2 & 4 & 30.69 & 30.41  \\ 
2 & 5,6 & 31.73 & 31.39 \\ \hline 
3 & 1   & 41.62 & 40.10 \\
3 & 2,3 & 44.39 & 42.40$^{\rm a}$ \\
3 & 4   & 44.39 & 42.42$^{\rm a}$ \\
3 & 5,6 & 45.23 & 44.40 \\
3 & 7   & 45.23 & 45.83 \\
3 & 8,9 & 50.07 & 49.36 \\
3 & 10  & 50.07 & 49.85 \\
\end{tabular}
\label{ENLEVEL}
\vspace{0.2 in} \noindent
a)  These energies are accidentally almost identical.  However,
group theory indicates that these levels are generically
nondegenerate.
\end{table}

\begin{table}
\caption[]{Matrix elements of $< N \alpha | A_{2}^{\tau}(r) | 0 1> $
(in meV) for  H$_2$ in octahedral and S$_6$ potential, respectively. 
Elements not listed are expected to be zero by symmetry. Numerically
such elements were found to be very small.  For this table the
wave functions  
within each degenerate manifold 
were chosen
to make the matrix elements of $A_2^\tau({\bf r})$ as simple
as possible.  For Pa${\bar 3}$ symmetry, the z-axis is taken
to be the local three-fold axis. Therefore the octahedral wavefunctions are
not necessarily identical to the Pa${\bar 3}$ wavefunctions.}
 
\begin{tabular}{ccrcc}  
\multicolumn{3}{c}{$< N \alpha | A_{2}^{\tau}({\bf r}) | 0 1 > $}
& Octahedral$^{\rm a}$ & Pa${\bar 3}$ (S$_{6}$) \\ \cline{1-3}
N &  $\alpha $ & $\tau $ & & \\ \hline  
0 & 1  & 0 & (0,0) & (--1.286,0) \\ 
2 & 1  & 0 & (0,0) & (--0.506,0) \\ 
2 & 1  & 1 & ($\alpha$,0) & (0,0) \\ 
2 & 1  & 2 & (0,0) & (0,0) \\ 
2 & 2  & 1 & (0,$\alpha$) & (--0.176,--0.106) \\ 
2 & 3  & 2 & (0,$\alpha$) & (2.413,--2.388) \\ 
2 & 4  & 0 & (0,0) & (--0.224,0) \\ 
2 & 5  & 0 & (0,0) & (0,0) \\ 
2 & 5  & 2 & ($\beta$,0) & (--6.500,--1.754) \\ 
2 & 6  & 0 & ($\sqrt 2 \beta$,0) & (0,0) \\
2 & 6  & 1 & (0,0) & (--8.862,--1.400) \\
\end{tabular}
\label{ALM0N}

\vspace{0.2 in} \noindent
a) Our numerical results give $\alpha=0.859$ and $\beta=5.662$ (in meV).
\end{table}

\begin{table}
\caption[]{Non-zero matrix elements of $<1\alpha | A_{2}^{\tau}(r) | 1 \beta> $
(in meV) for  H$_2$ in
octahedral and S$_6$ potential, respectively. We also note that
$<1\alpha | A_{2}^{\tau}(r) | 1 \beta> = <1\beta | A_{2}^{\tau}(r) | 1 \alpha> $
and $<1\alpha| A_{2}^{-\tau}(r) | 1 \beta> = (-1)^{\tau} <1\alpha| A_{2}^{\tau}(r) | 1 \beta>^{*}$.
For the octahedral symmetry,  the
wave functions within each degenerate manifold were chosen
to make the matrix elements of $A_2^\tau({\bf r})$ as simple
as possible.
}

\bigskip 
\begin{tabular}{ccrcc}  
\multicolumn{3}{c}{$<1\alpha | A_{2}^{\tau}(r) | 1 \beta> $}
& Octahedral & Pa${\bar 3}$ (S$_{6}$) \\ \cline{1-3} \\
$\tau$ & $\alpha $ & $\beta $ & & \\ \hline
0 & 1 & 1 & (--3.853,0) & (--2.020,0)       \\
0 & 2 & 2 & (--3.853,0) & (--1.326,0)       \\
0 & 3 & 3 & ( 7.708,0)  & (--1.326,0)       \\
1 & 1 & 2 & (0,0)       & (--2.109,--1.131) \\
1 & 1 & 3 & (0,0.881)   & (1.137,--2.104)   \\
1 & 2 & 2 & (0,0)       & (1.534,2.440)     \\
1 & 2 & 3 & (--0.881,0) & (2.435,--1.533)   \\  
1 & 3 & 3 & (0,0)       & (--1.531,--2.442) \\ \hline 
2 & 1 & 1 & (--4.119,0) & (0,0)             \\ 
2 & 1 & 2 & (0,0.880)   & (--0.225,--3.022) \\ 
2 & 1 & 3 & (0,0)       & (--3.022,0.226)   \\ 
2 & 2 & 2 & (0,0)       & (0.560,0.835)     \\ 
2 & 2 & 3 & (0,0)       & (--0.830,0.556)   \\ 
2 & 3 & 3 & (0,0)       & (--0.565,--0.831) \\
\end{tabular}
\label{A2MNNP}
\end{table}

\begin{table}
\caption{Shift of the center of gravity (CG) and  splitting
(in meV) of the $(J=1, N=0)$ manifold when the nominally octahedral
site has octahedral and S$_6$ symmetry.}
 
\begin{tabular}{|l | c c | c|}  
& \multicolumn{2} {c|} {Our Calculations} &
Experiment$^{\rm d}$ \\
Quantity & Octahedral (O$_{\rm h}$) & Pa${\bar 3}$ (S$_{6}$) & 
Pa${\bar 3}$ (S$_6$) \\ \hline
Shift of CG $^{\rm a}$ & -0.134 & $-0.141 $ & 0.35 \\ 
Shift of CG $^{\rm b}$ &  & $-0.16 $ & 0.35 \\ 
Splitting 1st order$^{\rm c}$   & 0 & 0.487 & \\ 
2nd order$^{\rm c}$   & 0 & -0.010  & \\ 
total$^{\rm c}$    & 0 & 0.477 & 0.70  \\ 
total$^{\rm b}$    & 0 & 0.46  & 0.70 \\ 
\end{tabular}
\vspace{0.2 in} \noindent
a) Perturbation result of Eq. (\ref{CGEQ}). 

\noindent
b) Obtained by direct diagonalization of
Eq. (\ref{HJEQ}) 

\noindent 
c) Perturbation result of Eq. (\ref{SPLITEQ}).  

\noindent 
d) From Ref. \onlinecite{NIST}.
\label{SPLIT}
\end{table}

\begin{table}
\caption{Energy (in meV) of $J=1$ single-phonon states where
zero-of-energy is taken to be 2B.}
 
\begin{tabular}{|c|cc|c|cc|}  
\multicolumn{3} {c} {Octahedral (O$_{\rm h}$)} &
\multicolumn{3} {|c} {Pa${\bar 3}$ (S$_{6}$)} \\ \hline
$\Psi$ & Full Mesh$^{\rm a}$ & Perturbation$^{\rm b}$ &
$\Psi$ & Full Mesh$^{\rm a}$ & Perturbation$^{\rm b}$ \\ \hline
$T_{1{\rm g}}$ & 13.11 & 13.14  & A$_{\rm g}$ & 12.59 & 12.52 \\
$T_{1{\rm g}}$ & 13.11 & 13.14  & E$_{\rm g}$ & 12.82 & 12.87 \\
$T_{1{\rm g}}$ & 13.11 & 13.14  & E$_{\rm g}$ & 12.82 & 12.87 \\
$T_{2{\rm g}}$ & 13.6 & 13.68  & E$_{\rm g}$ & 13.47 & 13.51 \\
$T_{2{\rm g}}$ & 13.6 & 13.68  & E$_{\rm g}$ & 13.47 & 13.51 \\
$T_{2{\rm g}}$ & 13.6 & 13.68  & A$_{\rm g}$ & 14.20 & 14.16 \\
$A_{\rm g}$ & 15.61 & 15.78  & A$_{\rm g}$ & 15.75 & 15.31 \\
$E_{\rm g}$ & 16.46. & 16.60  & E$_{\rm g}$ & 16.60 & 15.79 \\
$E_{\rm g}$ & 16.46. & 16.60  & E$_{\rm g}$ & 16.60 & 15.79 \\
\end{tabular}
\vspace{0.2 in} \noindent
a)  Solution to Eq. (\ref{HJEQ}) for the three component wavefunction
on a mesh of points.

b)  Solution to Eq. (\ref{HEFFEQ}) using wavefunctions and energies for
a $J=0$ molecule as previously determined numerically  on a mesh of points.
\label{JN1}
\end{table}

\begin{table}
\caption[]{Basis functions within the one phonon
$(J=1)$ manifold.  Here $x$, $y$, and $z$ are
the one phonon states with a single excitation in the phonon
associated with the $x$, $y$, and $z$ direction, respectively.$^{(a)}$
In terms of the $m_J$ states (denoted $|m_J\rangle$) within $(J=1)$
we have $X \equiv (|-1\rangle  - |1\rangle )/\sqrt 2 $, 
$Y \equiv i(|1 \rangle + |-1\rangle)/\sqrt 2$, and $Z \equiv |0\rangle$.}

\vspace{0.2 in}
\begin{tabular} {c | c }
\multicolumn{2} {c} {O$_{\rm h}$ Symmetry$^{\rm a}$} \\ \hline
T$_{2 \rm g}$ & $(xZ + zX), \ (yZ+zY), \ (xY+yX)$ \\ 
T$_{1 \rm g}$ & $(xZ - zX), \ (yZ-zY), \ (xY-yX)$ \\ 
E$_{\rm g}$ & $(2zZ -yY -xX) , \ (xX - yY)$ \\
A$_{\rm g}$ & $(xX + yY + zZ)$ \\
\hline
\multicolumn{2} {c} {Pa${\bar 3}$ Symmetry$^{\rm b}$} \\ \hline
E$_{\rm g}$ and E$_{\rm g}^*$ & $(xZ +zX , yZ+zY)\ ,
(xZ-zX, yZ-zY) , \ (xX-yY, xY+yX)$ \\
A$_{\rm g}$ & $zZ, \ xX+yY, \ xY-yX$ \\
\end{tabular}

\vspace{0.2 in}
a) The $x$, $y$, and $z$ directions are taken to coincide with the
four-fold axis of O$_{\rm h}$.

b) The $z$ direction coincides with the three-fold axis of S$_6$.
\label{SYM}
\end{table}


\newpage 

\newpage 

\newpage
\begin{figure}
\centerline{\psfig{figure=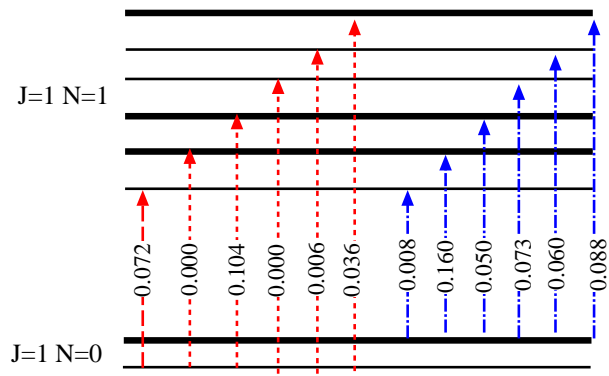,width=8cm}}
\vfill
\caption{The calculated transition probabilities from the
$(J=1,N=0)$ levels to the nine-fold manifold of $(J=1,N=1)$ levels
at T= 4~K.  For each pair of energies these transition probabilities
represent the appropriate sum over degenerate levels.
Note that there are at least eight transitions with
comparable probability, suggesting that rich features could 
be observed in a neutron scattering experiment.}
\label{PROBABILITIES}
\end{figure}

\newpage 
\begin{figure}
\centerline{\psfig{figure=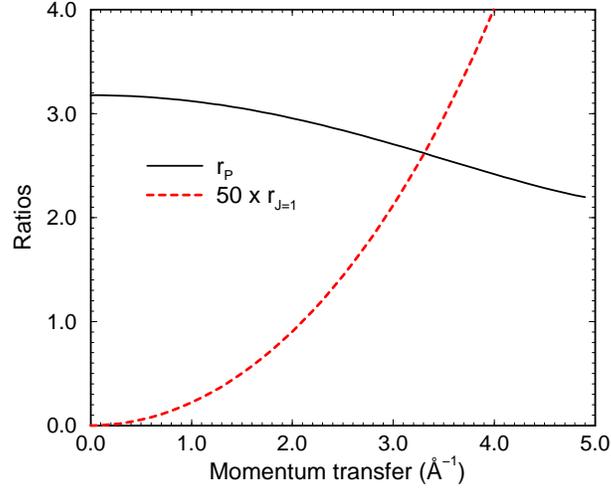,width=8cm}}
\vfill
\caption{The solid curve is the ratio 
$r_P$ of the single-phonon cross section to
that from para-ortho conversion as given in Eq. (\ref{OPr})
as a function of momentum transfer. The dotted line shows
$50 \times r_{J=1}$ (T=0) as given in Eq. (\ref{rj}).  The experimental 
situation of Ref. 5 corresponds to a momentum transfer
between 2 and 4 $\AA^{-1}$.}
\label{O-P}
\end{figure}

\newpage
\begin{figure}
\centerline{\psfig{figure=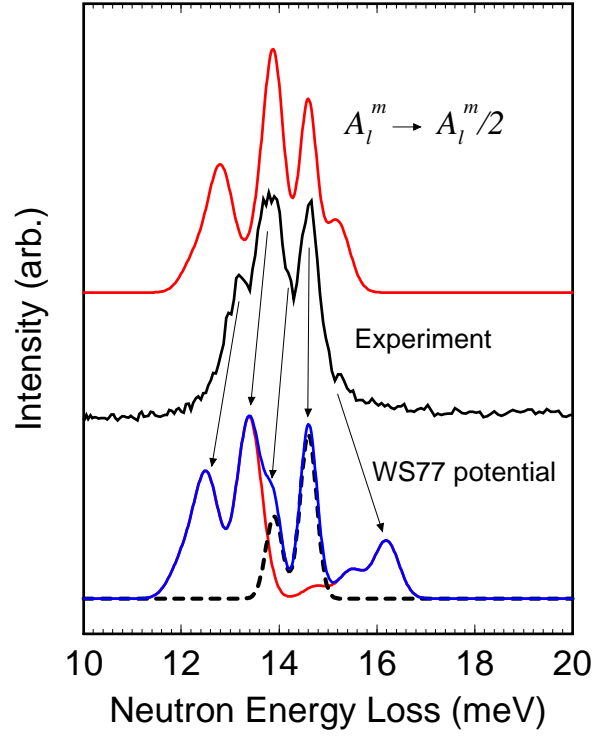,width=8cm}}

\vfill
\caption[]{Neutron energy loss spectrum (middle) at 4.2 K.
The bottom    curve is the result from our calculation
using the WS77 potential. The dashed and gray lines are
the contributions from rotational and vibrational excitations,
respectively. The top curve is the spectrum after arbitrarily
scaling $A_{l}^{m}$
by half, indicating that the orientational potential
used in our calculations is too anisotropic as
far as the center-of-mass motion is concerned.}
\label{expvscal} 
\end{figure}

\newpage
\begin{figure}
\centerline{\psfig{figure=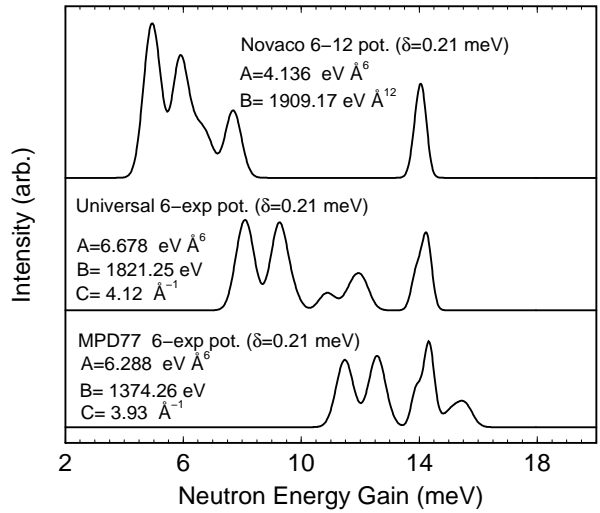,width=8cm}}

\vfill
\caption[]{Neutron energy loss spectrums obtained from various
commonly used intermolecular potentials. For each potential
we give the value of $\delta$, the
splitting of the $(J=1)$ zero-phonon levels, which may be
compared to the experimentally determined value 
$\delta=0.75$ meV.\cite{NIST} Note that the 
average phonon energies of the first two potentials (top and
middle curves) are way off from the observed value of $\approx 14$
meV. }
\label{otherpot} 
\end{figure}

\newpage
\begin{figure}
\centerline{\psfig{figure=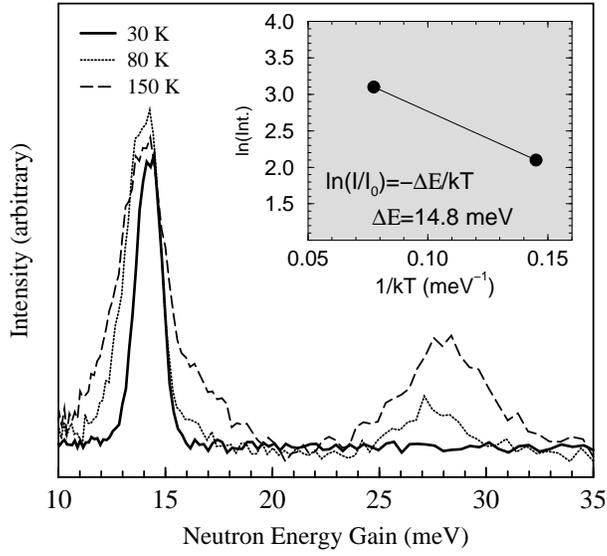,width=8cm}}

\vfill
\caption[]{Temperature dependent neutron energy gain spectrum
of H$_2$ in C$_{60}$ (the data is taken from Ref. \onlinecite{NIST}).
The inset shows  $\ln (I/I_{0})$ versus $1/kT$,
where $I$ is the intensity of the feature
at about 28 meV. The slope of the line indicates
an activation energy barrier of  14.8 meV.}
\label{28FIG} 
\end{figure}

\end{document}